\PassOptionsToPackage{hyphens}{url}

\documentclass[%
 reprint,
 groupedaddress,
 showpacs, 
 noshowkeys,
 amsmath,amssymb,
 aps,
 prl,floatfix
]{revtex4-1}
\usepackage{color}
\usepackage{graphicx}
\usepackage{dcolumn}
\usepackage{bm}
\usepackage{amsmath}

\usepackage[cdot,mediumqspace,amssymb]{SIunits} 
\usepackage[autolanguage]{numprint} 

\usepackage{ifthen}
\usepackage{ifpdf}
\ifpdf
    \graphicspath{{Figures/PDF/}}
\else
    \graphicspath{{Figures/EPS/}}
\fi

\usepackage{breakurl}
\begin{document}

\title{Measuring and Modeling Behavioral Decision Dynamics in Collective Evacuation}

\author{Jean M. Carlson$^{1}$, David L. Alderson$^{2}$, Sean P. Stromberg$^{1,*}$,  \\Danielle S. Bassett$^{1,3}$, Emily M. Craparo$^{2}$, Francisco Gutierrez-Villarreal$^{2}$, Thomas Otani$^{2}$} \affiliation{$^1$Department of Physics, University of
California, Santa Barbara, CA 93106, USA; $^2$Naval Postgraduate School,
Monterey, CA 93943; $^3$Sage Center for the Study of the Mind, University of
California, Santa Barbara, CA 93106} \email[Corresponding author. Email
address: ]{stromberg@physics.ucsb.edu}

\date{\today}

\begin{abstract}
\begin{center}{\bf Approved for public release; distribution is unlimited.}\end{center}
Identifying and quantifying factors influencing human decision making remains
an outstanding challenge, impacting the performance and predictability of
social and technological systems. In many cases, system failures are traced to
human factors including congestion, overload, miscommunication, and delays.
Here we report results of a behavioral network science experiment, targeting
decision making in a natural disaster. In each scenario, individuals are
faced with a forced ``go'' versus ``no go'' evacuation decision, based on
information available on competing broadcast and peer-to-peer sources.  In
this controlled setting, all actions and observations are recorded prior to
the decision, enabling development of a quantitative decision making model
that accounts for the disaster likelihood, severity, and temporal urgency, as
well as competition between networked individuals for limited emergency
resources. Individual differences in behavior within this social setting are
correlated with individual differences in inherent risk attitudes, as
measured by standard psychological assessments. Identification of robust methods for quantifying human decisions in the face of risk has
implications for policy in disasters and other threat scenarios.

\end{abstract}

\maketitle

\section*{Introduction}
The development of new communication technologies enables rapid information
dissemination and decision making among groups of individuals, but it also
creates new challenges in the coordination of collective behavior. For
example, the adoption of social networking technologies such as Twitter and
Facebook can empower the masses but makes them hard to
control~\cite{oliver_theory_1985,gonzalez-bailon_dynamics_2011,Howard2011,chen2011,Khondker2011,LotanEtAl2011,Farrell2012,Barnsby2012}.
More generally, the advent of contemporary network technologies has brought
with it a new set of fragilities stemming from the complexity of human
behavior: people rarely behave optimally, randomly, or uniformly, as often
naively assumed in technological design and policy development.

Within the field of network science, the study of social networks plays an
increasingly important role in method development and associated
applications, with widespread implications in marketing, politics, education,
epidemics, and disasters. Considerable effort is directed towards understanding
how information diffuses through social groups
\cite{WeakTies-PRE-2010,GomezEtAl-PRL-2013,MyersEtAl-arxiv-2012,GuilleEtAl-arxiv-2013,social-cascades-arxiv-2013,info-cascades-arxiv-2013},
with particular emphasis on the role of news websites \cite{Leskovec2009},
blogs \cite{Leskovec2007},  Facebook \cite{Onnela2010}, Twitter
\cite{Bakshy2011}, and other social media \cite{LermanGhosh2010,Simmons2011}.

As information diffuses, individuals can display a range of decision making
behaviors driven by new information. Phenomena of particular interest include
(1) the dynamics of cascading behavior, which can explain how and why fads
emerge \cite{Watts2002} or rumors spread so quickly
\cite{Doerr2012,ZhangEtAl-PRE-2013}, and (2) the role that individuals play
as ``spreaders'' in facilitating the propagation of this behavior
\cite{SpreadersNature2010,SpreaderAbsence-PRE-2012,FindingSpreaders-PRE-2012}, or
similarly the roll that ``homophily'' can play in abrogating uptake of a 
behavior \cite{Centola2011}.
Social epidemics, much like their biological counterparts
\cite{Dodds2005,Bettencourt2006,epidemic-intelligence-arxiv-2012,epidemic-threshold-arxiv-2013},
are often modeled as single- \cite{Burt1987} or multi-stage \cite{Melnik2011}
complex contagion processes \cite{Centola2007,Centola2007b,Centola2010}.

Recent theoretical investigations have examined how this information exchange
leads to collective action. In one class of models, individual agents occupy
nodes on a network, and a set of rules defines information propagation
dynamics and individual decision making behavior (e.g., see
\cite{Dodds2005,BassettEtAl2012-PRE,ZhangEtAl-PRE-2013}). Complementary data
driven investigations describe computational algorithms that begin to unravel
rules for influence and decision making from large databases, such as
Twitter, Facebook, and wireless communication networks (e.g.,
\cite{LotanEtAl2011,BarahonaEtAl2012,FindingSpreaders-PRE-2012,SanoEtAl-PRE2013}).
In most cases the databases identify decisions that are made and delineate
links between network members. However, information about the factors that
drive human decisions, including individual observations, attention, history,
personality, and risk perception is generally unavailable.

This paper focuses on a critical link between simulation studies and
empirical observations of large scale networks. Specifically, we conducted a
behavioral experiment involving a group of 50 individuals in a computer
laboratory. Because human behavior is often far from what is predicted by
idealized models, experimental observation in ``live'' and controlled
environments are essential for improved understanding and modeling of social
phenomena. Our work adapts the framework of Kearns \textit{et al.}~\cite{Kearns2006,Kearns2009,Kearns2010,Kearns2012}, who have conducted a
series of ``behavioral network science'' (BNS) experiments that have focused
on collective problem solving tasks, such as abstract graph coloring problems
or economic investment games. These experiments, and similar experiments from other research groups, have demonstrated that
``human subjects perform remarkably well at the collective level'' in a
number of tasks and scenarios, both competitive and cooperative
\cite{Kearns2012, mason_collaborative_2012, nedic_decision_2012}. However, disasters and other crisis situations often
display the opposite
effect~\cite{Drabek1986,LPP06,DH07,Sweeney2008,Conneally2011}. Social
interactions affect traffic flow \cite{helbing_simulating_2000, helbing_traffic_2001}, and can lead to a ``mob mentality'' \cite{Banerjee1992,Bosse2012,Edelson2011} that hinders evacuation and may
lead to injury and violence. Moreover, associated spatiotemporal clustering of
departure times can lead to traffic congestion and
delays~\cite{CS02,DOT06,H11}.

Therefore, in contrast to previous BNS research involving idealized, abstract
games, our investigations involve decision making in a threat scenario.
Specifically, our study is set in the context of an impending natural
disaster, where each individual occupies a node in a social network and must
decide whether or not to evacuate. The experiment is conducted for a sequence
of time-evolving disaster scenarios. In each scenario, individuals receive
real time updates from a centralized information source about the likelihood,
severity, and timing of a disaster that threatens their virtual community.
Individuals also receive social information regarding decisions of their
neighbors, and availability of space in a virtual shelter. Thus, participants
face a tradeoff in competing types of information (i.e., centralized
broadcast information versus decentralized social information) in a
laboratory setting that emphasizes risk and loss.

Compared to large data driven studies, the experiment provides a much more
complete, quantitative set of measurements, enabling us to assess factors and
isolate tensions that arise in human decision making. In addition to
observing the ultimate evacuation decisions, our experimental setup allows us
to monitor the behavior of individuals as they gather information. Prior to
the experiment, we also assess individual personality profiles and risk
attitudes using standardized tests. The ability to acquire this extensive set
of static and dynamic measurements both prior to and during the
decision making process allows us to quantify links between psychological
assessments and heterogeneity in group behavior.

A primary outcome of this study is the identification of a decision model for
evacuation behavior based on empirical observations. The model output fits
the observations remarkably well and can be used to quantify individual
differences in decision dynamics. The empirical model reduces the catalog of
scenarios and observations to a few key parameters involving an overall multiplicative rate factor for evacuation, an average decision threshold based on the disaster likelihood, and variability about the average threshold, reflecting how consistently the decision making threshold was applied.
The model enables us to isolate and compare two sources of urgency in the
experiment that differentially impact observed behavior: time pressure for the
evacuation decision and competition for shelter space. This empirical model
stands in contrast to a set of models typically used in numerical simulations
or large scale, data driven studies that treat decisions as random, optimal,
or based on a threshold applied to a state variable representing opinion,
which is updated by an assumed interaction rule (e.g.,
\cite{Banerjee1992, Watts2002, Sood2008, H11, BassettEtAl2012-PRE, Bosse2012, helbing_simulating_2000, helbing_traffic_2001}).

While our experiment is admittedly well removed from a true natural disaster,
it allows us to isolate and quantify tensions that arise in a crisis, in a
manner that would not be possible during an actual event. Furthermore, the
experimental design takes into account known psychological factors associated
with risk perception, threat, and information processing
\cite{Gigerenzer2002,Chib2012,Edwards2012,BD12}. A key component of
behavioral network science is to use the observed human behavior as
inspiration for the development of novel computational models of behavior,
which can in turn be tested in future experiments.  This spiral development
of {\it model-experiment-model} or {\it experiment-model-experiment} may be
used to develop optimal strategies for disseminating information during a
disaster, and insuring sufficient allocation of resources for disaster
response.

\section*{Materials and Methods}
On May 18, 2012 an experiment was conducted at the University of California,
Santa Barbara (UCSB) in which 50 student participants within a virtual
community each decided if and when to evacuate from impending natural
disasters. All participants provided written informed consent, and the
experimental protocol was approved by the Institutional Review Board of UCSB.
The demographic composition of the participants was not released for publication. 
Prior to taking part in the study, the personality profile of each
participant was measured using the Big Five Inventory (BFI-44) questionnaire
\cite{John2008,John1991,Benet1998}, and the risk preferences of each
participant were also measured in six domains (social, investment, gambling,
health \& safety, ethical, and recreational) using a Domain Specific
Risk Attitude Scale \cite{Weber2002,Blais2006}. The Big Five Inventory is a
commonly used set of 44 questions that enables the assessment of an
individual's personality along the following dimensions: extraversion,
neuroticism, openness, conscientiousness, and agreeableness. The Big Five is
used extensively in psychological research as well as in translational
applications such as the assessment of learning styles and employee
placement. The Domain Specific Risk Attitude Scale is used in psychological
research to assess risk perception and risk behavior, to predict human
behavior, and to develop policy in areas such as health and natural hazards.
Administration of each questionnaire lasted approximately 7 minutes.

Individuals participated in 47 scenarios (runs) that lasted one minute each.
At the beginning of each scenario, each participant was given 100 monetary
``points'' that were at risk from a simulated disaster. During each scenario,
participants were provided with information about the progression of the disaster, and they were offered the opportunity
to evacuate from this disaster (a
binding decision) and occupy one of a limited number of spaces in a virtual
disaster shelter. Depending on their decision and the outcome of the
disaster, they could lose some or all of their monetary points. The magnitude
of the loss was a function of whether or not the individual successfully
evacuated to the shelter, and whether or not the disaster struck. The total
amount paid to a participant at the end of the experiment was a function of
their cumulative score over the 47 runs. The running cumulative scores of all of the participants were ranked and displayed on a leader board at the front of the room. This allowed 
individuals evaluate their strategy and provided a competitive 
incentive.

\subsection*{Experiment Layout}

The primary objective of this project was to understand the way in which
individual decision makers use and share information, and how this
information leads to collective action of the group as a whole. Of particular
interest was obtaining insight into the influence of competing sources of
information on individual and group behavior.

\begin{figure}[htbp]
\begin{center}
 \includegraphics[scale=1.0]{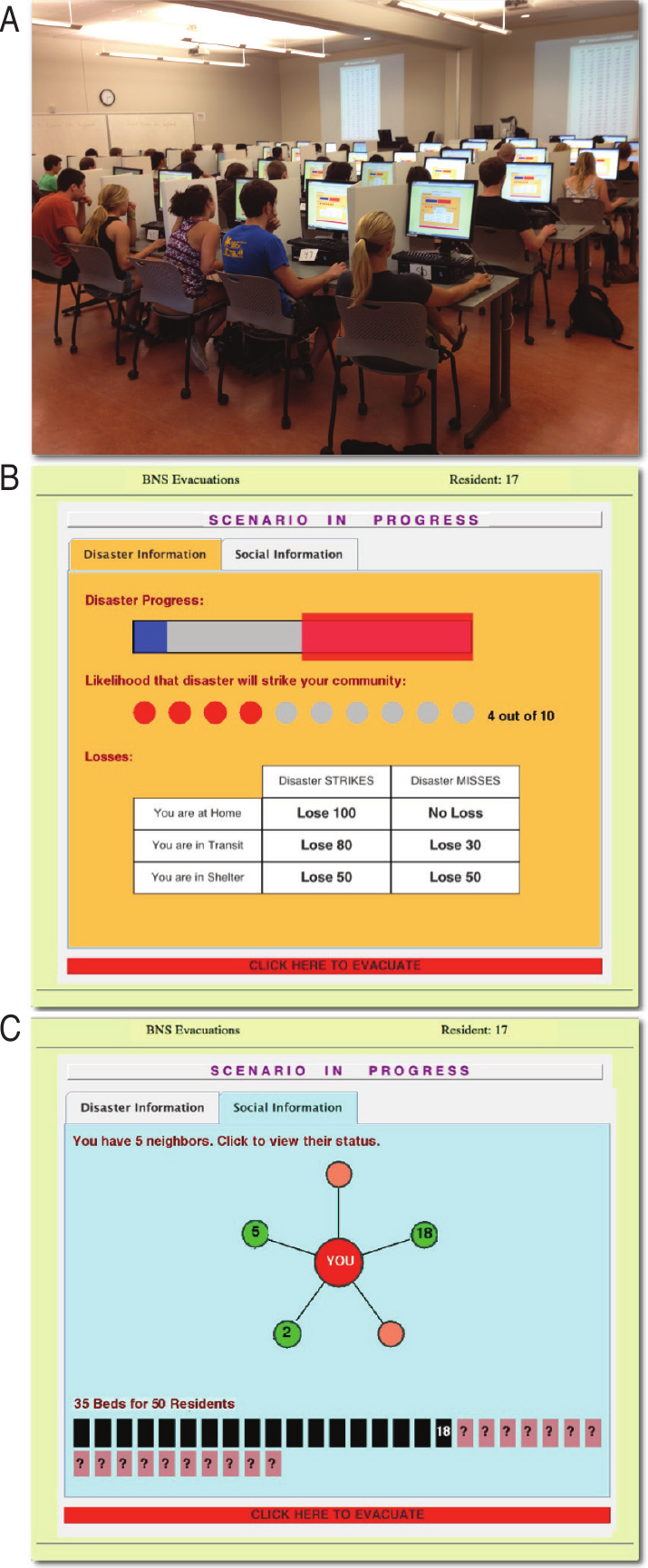}
\end{center}
\caption{{\bf \textbf{A}: Experimental setup at UCSB.}  \textbf{B}: Disaster Tab, showing current status and loss table.  \textbf{C}: Social Tab, showing status of neighbors; in this example, neighbors have claimed shelter spaces 2, 5, and 18, meaning that at least 18 of 35 shelter spaces have already been filled.}\label{tab1}
\end{figure}

To reach these objectives, we employ an experimental setup derived from that
of Kearns \textit{et al.}~\cite{Kearns2006,Kearns2009,Kearns2010,Kearns2012}. We
customize the computational framework and user interface to our evacuation
problem. Each participant sits in front of a computer screen, see
Figure~\ref{tab1}\textbf{A}, containing two tabbed windows, labeled
``Disaster Information'' and ``Social Information.'' The participant may only view one
window at a time and can switch between these two sources of information by clicking on the tabs.

The Disaster Tab, shown in Figure \ref{tab1}\textbf{B}, provides participants
with information about the simulated time-evolving disaster. At the top of
this tab is a disaster progress bar, which incrementally turns blue as time
goes by; a red box around the scenario progress bar signifies the time window
in which the disaster could strike. The likelihood that the evolving disaster will
strike the community is presented in real time as the proportion of filled circles (e.g.,
4 out of 10 filled circles indicates a current probability of 40\%).
A loss matrix shows how
many points an individual will lose at the end of the current scenario
depending on the outcome of the disaster and the individual's final location.
Finally, a button at the bottom of the
Disaster Tab allows participants to evacuate. When an individual clicks the
button, they transition from being ``AtHome'' to being ``InTransit.''  If
there is still space available in the shelter, the individual immediately
transitions to being ``InShelter.'' However, if the shelter is already full,
the participant remains InTransit through the rest of the current
scenario.

The Social Tab, shown in Figure \ref{tab1}\textbf{C}, allows the participant
to query the status of neighbors in their social network by clicking on each 
neighbor's node. If the neighbor is
still AtHome, then the letter `H' appears on the neighbor node. If the neighbor
has evacuated, a subsequent click is required to identify this. If the
neighbor is InTransit, then the letter `T' appears. If the neighbor is in
the shelter, then the shelter space (or ``bed'') number that the neighbor
occupies in the shelter appears. This value provides a lower bound on the
number of beds occupied in the shelter and is also recorded in a shelter
diagram toward the bottom of the Social Tab. The evacuation button located on
the Disaster Tab is mirrored on the Social Tab to enable participants to make
their evacuation decision irrespective of their current tab location.

\subsection*{Psychometrics of Participants}
\begin{figure}[htbp]
\begin{center}
	\includegraphics[scale=1]{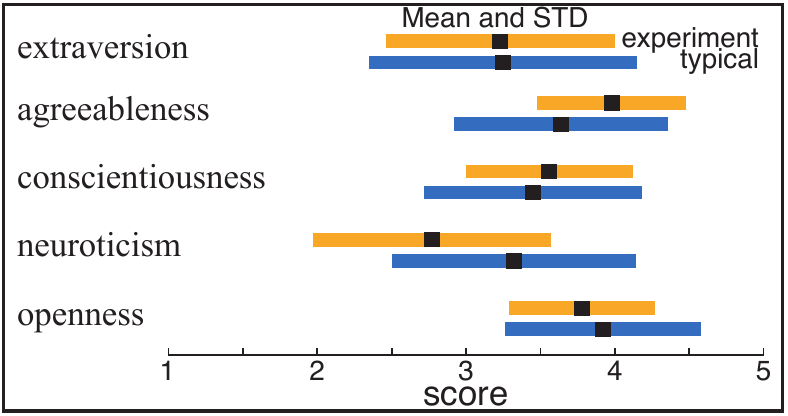}
\end{center}
\caption{{ \bf Mean and standard deviation (STD) for the Big Five Inventory scores} calculated over all 50 participants (yellow). For comparison, we report the typical values estimated from 6076 individuals aged 21 (blue) \cite{Srivastava2003}. The only significant deviation from typical scores was neuroticism, which had a significantly lower mean value.
\label{Tab2}}
\end{figure}

\paragraph{Personality metrics.} The Big Five Inventory measures an individual's personality based on five
characteristics: extraversion, agreeableness, conscientiousness, neuroticism,
and openness \cite{John2008,John1991,Benet1998}. As shown in Fig.~\ref{Tab2}, 
the group of individuals that volunteered to take part in our
experiment displayed similar personality profiles to the typical values for a
similar age group \cite{Srivastava2003}, with the exception of neuroticism
which was significantly lower than in the general population.

\paragraph{Risk Attitude:}
The risk attitude questionnaire scores both general risk attitude and
specific risk types in the following domains: investment, health \& safety,
gambling, social, ethical, and recreational. The evacuation scenarios in this
experiment were developed predicated on the assumption that individuals would
be averse to the loss of monetary points (financial risk), and loss of life
and property (health \& safety risk). Participant responses to questions on
the Domain Specific Risk Attitude Scale test ranged from ``1'' (Risk Averse) to ``5''
(Risk Seeking) with ``3'' indicating a risk neutral attitude. The general
risk attitude distribution was risk averse ($2.60\pm0.69$). When segregated
into the separate domains, the population displayed a range of risk attitudes
summarized in Table \ref{Tab1}.

\begin{table}[!ht]
	\caption{\bf{Risk Attitudes}}
	\begin{center}
\begin{tabular}{| l | l | l | l |}
\hline
Domain              & Mean  & STD   & Attitude Tendency \\
\hline \hline
Social              & 3.49  & 0.57  & Risk Seeking \\
Recreational        & 3.09  & 0.90  & Risk Neutral \\
Gambling            & 1.59  & 0.77  & Risk Averse \\
Health \& Safety    & 2.65  & 0.64  & Risk Averse \\
Ethical             & 2.02  & 0.56  & Risk Averse \\
Investment          & 2.76  & 0.92  & Risk Neutral \\
\hline
\end{tabular}
\begin{flushleft}{Risk attitude scores in 6 domains: mean and standard deviation (STD) calculated over all 50 participants.}
\end{flushleft}
\label{Tab1}
\end{center}
\end{table}

\subsection*{Scenario Simulation Mechanics}

Our experimental setup had several key features designed to enable the
isolation of external drivers and the identification of tradeoffs in
decision mechanics. These features included a network structure linking
participants and constraining information diffusion, time-evolving disaster
trajectories, and scenario-to-scenario variation in shelter capacity, time
pressure, and potential risk to monetary ``points''. We describe these
features in greater detail below.

\paragraph{Network Structure.}
In our experiment, a network structure enables participants to observe the actions
of others. In each scenario, participants are assigned at random to a node in
an underlying social network topology designed by the researchers. This
allows  an individual to have a different number of neighbors in each
scenario, and for the number of neighbors to vary by individual in a single scenario.
There were 8 networks used in the experiment: 3 ``regular'' ring lattice graphs,
where each node was connected to nodes within a distance 1, 2, or 3,
resulting in fixed node degree $d=$ 2, 4, or 6, respectively;
and 5 ``variable'' graphs where nodes
had degree $d \in [1,10]$ with an average $d=4$.
More specifically, the latter networks were generated
as random graphs with specified degree sequence
\{1($\times$10), 2($\times$8), 3($\times$7), 4($\times$6), 5($\times$5), 6($\times$4), 7($\times$4), 8($\times$3), 9($\times$2), 10($\times$1)\},
according to the algorithm specified in \cite{BayatiEtAl2010} and implemented in the {\tt NetworkX}
Python library \cite{hagberg-2008-exploring}. Number of neighbors was varied to measure the affect on frequency of seeking social information. Different network structures were used as they predict different
rates of information diffusion, with random networks having rapid diffusion, and 
regular lattice graphs having a slow rate of diffusion \cite{Easley2010}.

\paragraph{Disaster Trajectories.}
The disaster strike probability as a function of time $t$, denoted by
$P_\mathrm{hit}(t)$, was generated in advance from a well-defined stochastic
process (details of its construction can be found in \cite{CrewsThesis}).
The process corresponds to a two-dimensional
progression of a threat that moves toward a notional ``target'' with random
lateral motion in one dimension and monotonic forward progression in the
other dimension.  The lateral motion is simulated with a range of step sizes
limited by a prescribed volatility, while the forward motion may either have
variation or step deterministically. We record a ``Hit'' (corresponding to a
disaster strike) if the threat contacts a target, or a ``Miss'' if the
forward motion causes the threat to pass the target without hitting.
Participants can observe  a truncated value of $P_\mathrm{hit}(t)$ on the
Disaster Tab which is updated every second, however the overall trajectory is
not shown. There were a total of 23 $P_\mathrm{hit}(t)$ trajectories used in
the experiment, with many of the trajectories repeated with different
settings for other experimental variables.

\paragraph{Shelter Capacity.}
Scenarios varied in shelter capacity. There were 5 different shelter capacity
scenarios: 50, 40, 30, 20, and 10 beds. When the number of beds in the
scenario was less than 50 (the number of participants), individuals had to
compete for access to these beds and could access information on the
availability of shelter space through their social network.

\paragraph{Time Pressure.}
Scenarios varied in time pressure for an evacuation decision. When forward
motion in the disaster trajectory model was deterministic, the disaster would
either Hit or Miss at exactly 60 seconds. This type of time pressure is denoted ``CertainTime''. 
For runs with variable time
steps in the disaster trajectory model, the disaster could hit at any point
between 30 and 60 seconds, with an end time that is not known in advance to the
participants. We refer to this type of time pressure as ``VariableTime''. 
The distinction between these types of scenarios could be observed by participants through the
red box around the scenario progress bar on the Disaster Tab.
These different scenarios were designed to test how temporal
uncertainty affected evacuation strategies.

\paragraph{Potential Loss.}
Scenarios varied in potential risk to monetary ``points''. At the start of a
scenario, each participant is staked 100 points. The amount lost due to the
disaster depends on the loss matrix, the outcome of the scenario, and by the
individual's location at the end of the run (AtHome, InShelter, or InTransit).
Three loss matrices were used in the experiment and were based on underlying
incentive structures designed by the researchers, with the values changing
between runs acting to simulate varying disaster severity. The six entries in
the loss matrix (seen on the Disaster Tab) correspond to the combination of
the three end-state possibilities and the two disaster outcome possibilities.
All loss matrices had a 0 point loss for an (AtHome, Miss) outcome, with increasing loss for (InTransit, Miss) and (InShelter, Miss). When the disaster hit, loss is minimized for the combination (InShelter, Hit), followed by (InTransit, Hit), and the most costly outcome is (AtHome, Hit). While one could envision many disaster scenarios where it would be more costly to be InTransit than AtHome, our modeling choice was motivated by InTransit resulting in distancing oneself from the disaster epicenter, and more generally, taking some action rather than none. Values in the loss matrix were deliberately chosen to prevent trivial solutions, such as always evacuate or always stay home, from being winning strategies.

\begin{figure}[htbp]
\begin{center}
	\includegraphics[scale=1]{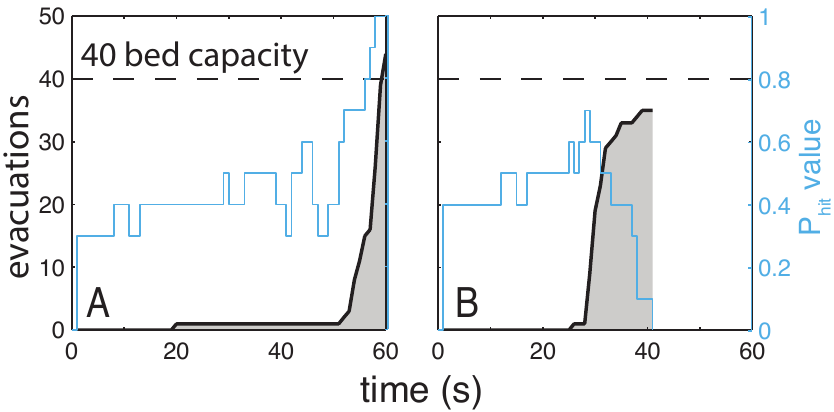}
\end{center}
	\caption{{\bf The collective evacuation behavior in two different scenarios.}  \textbf{A} (CertainTime): Participants wait until the end of the run to evacuate, waiting for more accurate information on the likelihood that the disaster will strike; some get stranded InTransit when the number of evacuees exceeds the shelter capacity.  \textbf{B} (VariableTime): More than half the participants evacuate at approximately the 30 second mark, which is the first time that the scenario could end. }
	\label{fig:Trajectories}
\end{figure}

\vspace{1ex}

To summarize our setup and participant behavior, we plot the cumulative
behavior for two evacuation scenarios in Figure \ref{fig:Trajectories}.  
The overall behavior in each scenario
can be observed by the interaction of the $P_\mathrm{hit}(t)$ trajectory (in
blue), the cumulative number of evacuations (grey fill), the number of
available shelter spaces (dashed line), and the end time of the scenario. The 
scenario in Figure \ref{fig:Trajectories}\textbf{A} is CertainTime
while the scenario in Figure \ref{fig:Trajectories}\textbf{B} is VariableTime.  In both scenarios, there are 40 shelter
spaces (beds) available for the 50 participants. In Figure
\ref{fig:Trajectories}\textbf{A}, we observe evidence of a stampede in which
participants evacuated for limited shelter space toward the end of the
scenario; some participants were left stranded in the state InTransit.
In Figure \ref{fig:Trajectories}\textbf{B}, we observe that a large number of
participants evacuated at approximately the 30 second point in the scenario
(the first time the run might end), but that the disaster did not happen.

\section*{Results}
The data collected during the experiment include every mouse click, for all
50 participants in each of the 47 disaster scenarios. From the data we can
identify what each individual was seeing, when they were seeing it, and if
and when they evacuated. This section describes empirical observations and
statistical analysis based on these results, which is used to develop a
quantitative decision model in the next section. Key variables include the
strike probability ($P_\mathrm{hit}$) trajectory (Fig.~\ref{fig:Trajectories}
blue), the loss matrix, the number of beds in the shelter
(Fig.~\ref{fig:Trajectories} dashed-black), and time pressure for the
evacuation decision.

\begin{figure}[htbp]
\begin{center}
	\includegraphics[scale=1.0]{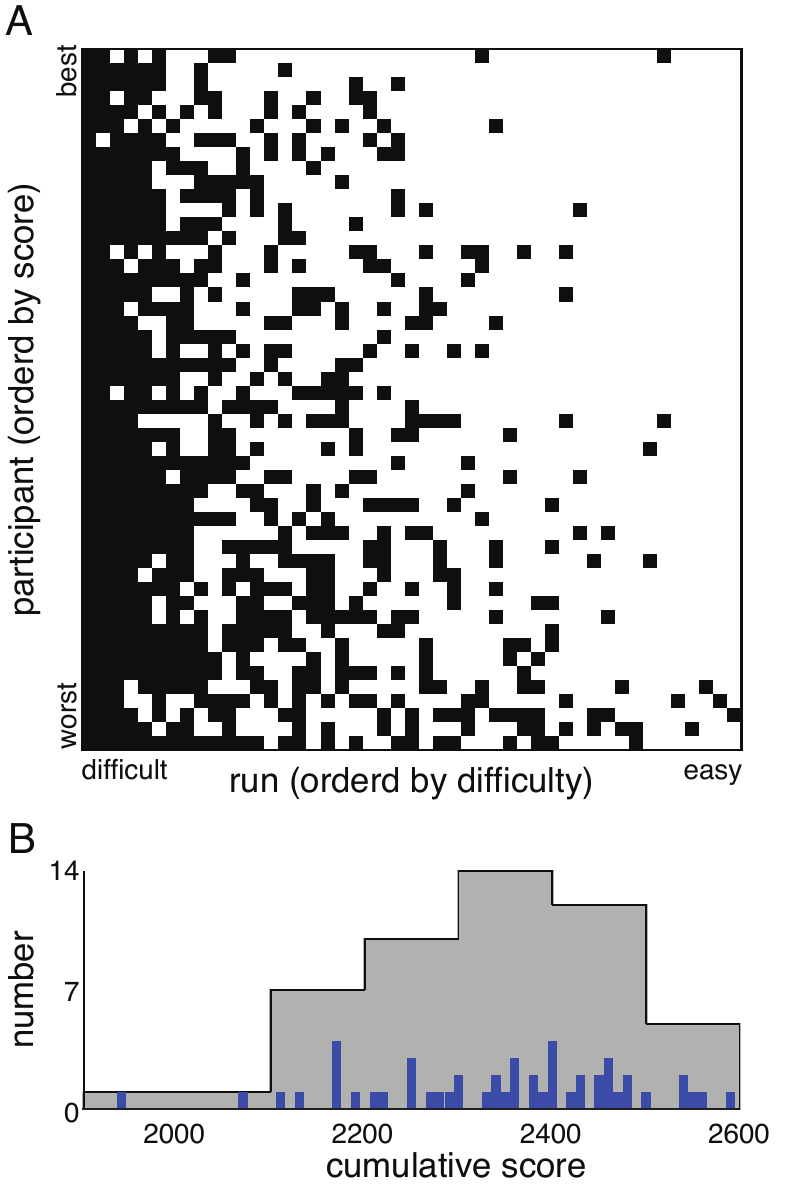}
\end{center}
	\caption{{\bf Success and distribution of cumulative scores.} \textbf{A} shows successful decisions in white [(InShelter, Hit); (InTransit, Hit); (AtHome,Miss)] and unsuccessful decisions in black. The participants are ordered by cumulative score, with the highest scoring at the top. The runs are reordered with the most difficult run on the left. \textbf{B} presents a histogram of the cumulative scores (grey), with bars showing the exact scores in blue. The blue bars highlight the divergence of the most unsuccessful participant.}
	\label{playerpayoffs}
\end{figure}

\paragraph{Participant rankings and scores.}
The success of each participant in each scenario is depicted in Figure
\ref{playerpayoffs}\textbf{A}. We quantify a participant's success using the
total point score retained at the conclusion of the 47 runs. The three types
of successful decisions [(InShelter, Hit); (InTransit, Hit); (AtHome,
Miss)] are shown in white, while unsuccessful decisions are shown in black.
In the ``hardest'' scenario (located towards the left-most side of the panel
in Figure \ref{playerpayoffs}\textbf{A}), there were zero successes in the
population, while in the ``easiest'' scenarios (located towards the
right-most side of the panel) a single participant was unsuccessful in each
run.

The distribution of cumulative scores is skewed: the lowest scoring
participant is far below the rest (see Figure \ref{playerpayoffs}\textbf{B}).
We analyze the differences in decision making patterns for different individuals
in more detail in a later section entitled Individual Variation.

\paragraph{Participant score correlates with risk attitude.}
We hypothesized that risk attitude in both the financial domain and the health \& safety
domain would be a significant factor in overall performance. We estimated an
individual's general financial risk attitude by averaging their scores from
both gambling and investment risk domains \cite{Blais2006,Weber2002}, and we
estimated their overall performance using the cumulative score. Cumulative
score was significantly correlated with health \& safety risk attitude
(Pearson correlation: $r=-0.31$, $p=0.02$) but not with financial risk attitude ($r=-0.04$, $p=0.73$). These results indicate that individuals that
were more averse to health \& safety risks (and therefore potentially more
susceptible to the specific influences associated with an evacuation decision scenario) performed
better than those that were less averse.

\begin{figure*}[htbp]
\begin{center}
	\includegraphics[scale=0.75]{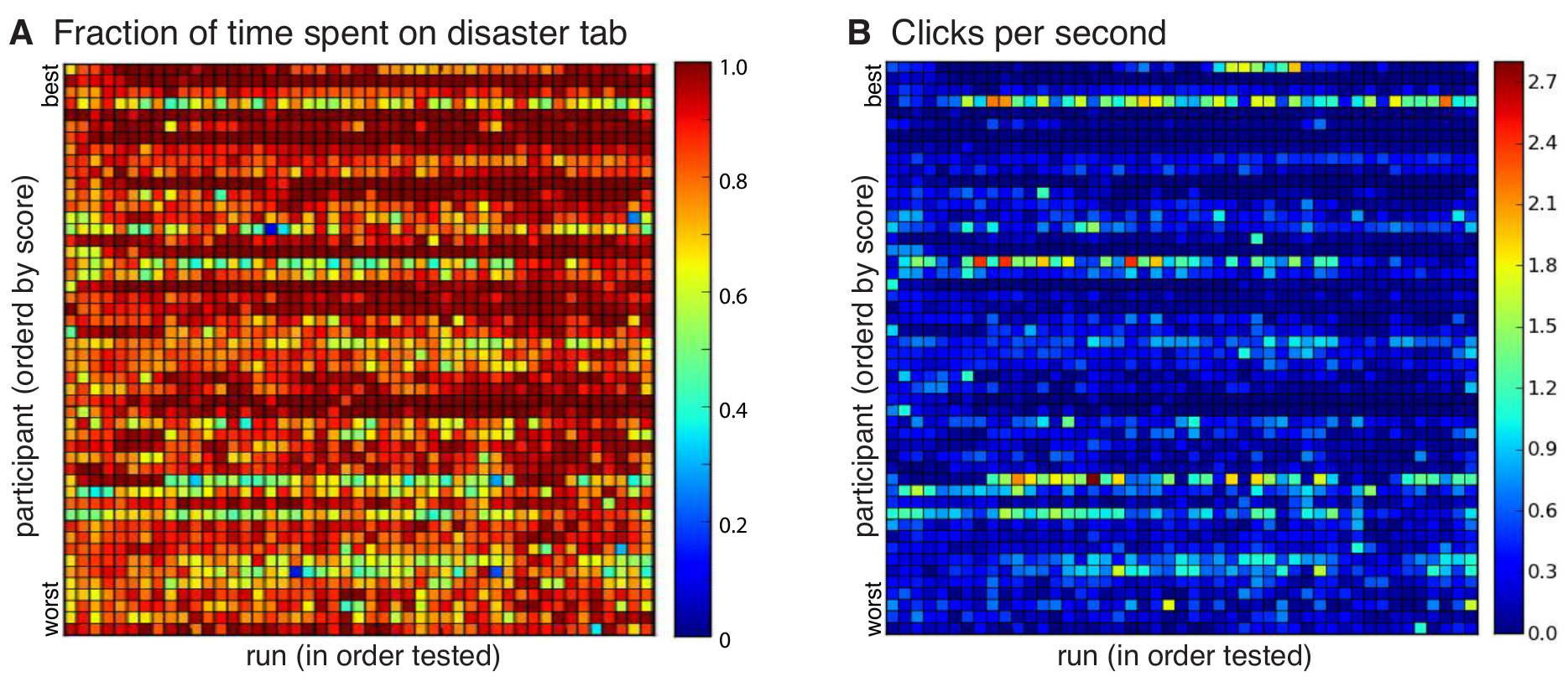}
\end{center}
	\caption{{\bf Participants spent the majority of their time on the Disaster Tab} (Frame \textbf{A}), but we can see those who spent more time on the Social Tab also had higher click frequency (Frame \textbf{B}) likely the result of trying to gain information on remaining shelter space.}
	\label{clicksperformance}
\end{figure*}

An interesting question is whether the observed correlation between risk attitude and performance was consistently observed over the population or
whether it was driven by a subset of individuals. From a psychological
perspective, one meaningful segregation of individuals into groups is a
partition based on the consistency of individual risk preferences across
domains. Individuals with consistent risk preferences across domains often
display different personality traits --- which could directly lead to
differences in behavior --- than those with inconsistent risk preferences
across domains \cite{Soane2005}. To estimate the consistency of risk attitudes we computed the standard deviation $\sigma$ of mean scores across
the 6 risk domains. We separated participants into a ``consistent'' group,
composed of those individuals with $\sigma<1$ (N=31), and an ``inconsistent''
group, composed of those individuals with $\sigma>1$ (N=19). The observed
correlation between performance and health \& safety risk attitude appears to
be driven by individuals with inconsistent risk attitudes ($r=-0.50$,
$p=0.02$) rather than by individual with consistent risk attitudes
($r=-0.18$, $p=0.32$). This suggests that individuals with domain specific
risk attitudes might tune their behavior more closely to the risk structure
of the experiment.

\paragraph{Participants focus on Disaster Tab.}
Our results indicate that participants viewed
the Disaster Tab more than the Social Tab. Individuals spent the vast
majority of their overall scenario time on the Disaster Tab, and they made
99\% of evacuation decisions while on this tab (see
Fig.~\ref{clicksperformance}\textbf{A}). Although on average participants did not
tend to spend as much time on the Social Tab, there was significant variation.
We did not observe a significant relationship between time spent on each tab
and performance.

\paragraph{Clicking behavior links to Social Tab.}
Click frequencies for all participants in all scenarios are shown in
Figure~\ref{clicksperformance}\textbf{B}, which lists participants by their
overall performance (highest first). We can see from this figure that the
higher click frequency individuals spent less time on the Disaster Tab and
therefore more time on the Social Tab. The majority of participants displayed
low values of clicking activity, indicating that they accessed social network
information infrequently. We did not observe a significant relationship
between click frequency and performance.

\begin{figure}[htbp]
\begin{center}
	\includegraphics[scale=1]{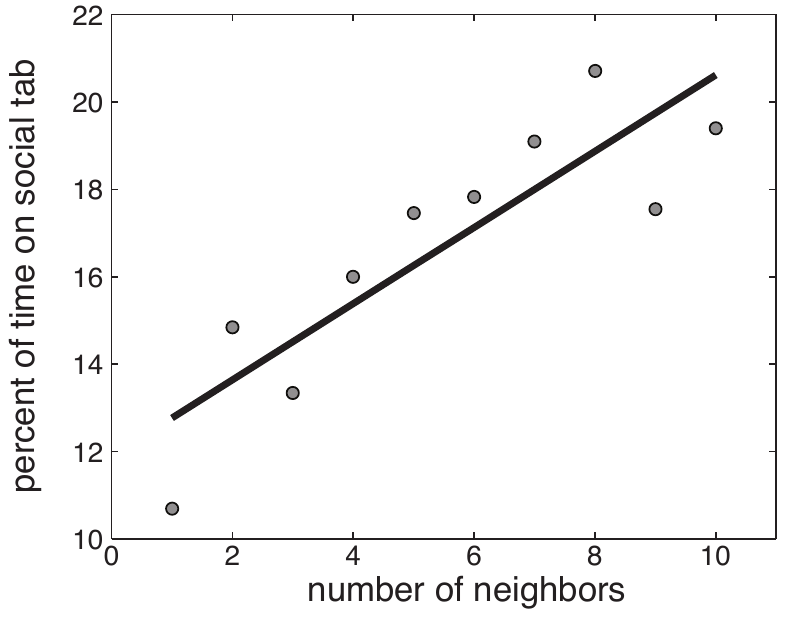}
\end{center}
\caption{{\bf Relationship Between Number of Neighbors and Time Spent on Social Tab.} The more network connections a participant had, the more time they spent on the social tab, with a Pearson correlation $r= 0.8690$, $p = 0.0011$. }
	\label{Fig_socialtime}
\end{figure}

\paragraph{Network structure drives time spent on Social Tab.}
The total number of neighbors a participant could have in any single scenario
ranged between one and ten. Fig.~\ref{Fig_socialtime} shows that participants
with many neighbors tended to spend more time on the Social Tab than those
with few neighbors. This result is intuitively consistent with the fact that
highly connected individuals could gain more social information than less
connected individuals, and might therefore be predisposed to spend more time
on the Social Tab to obtain this information.

\begin{figure*}[htbp]
\begin{center}
	\includegraphics[scale=1]{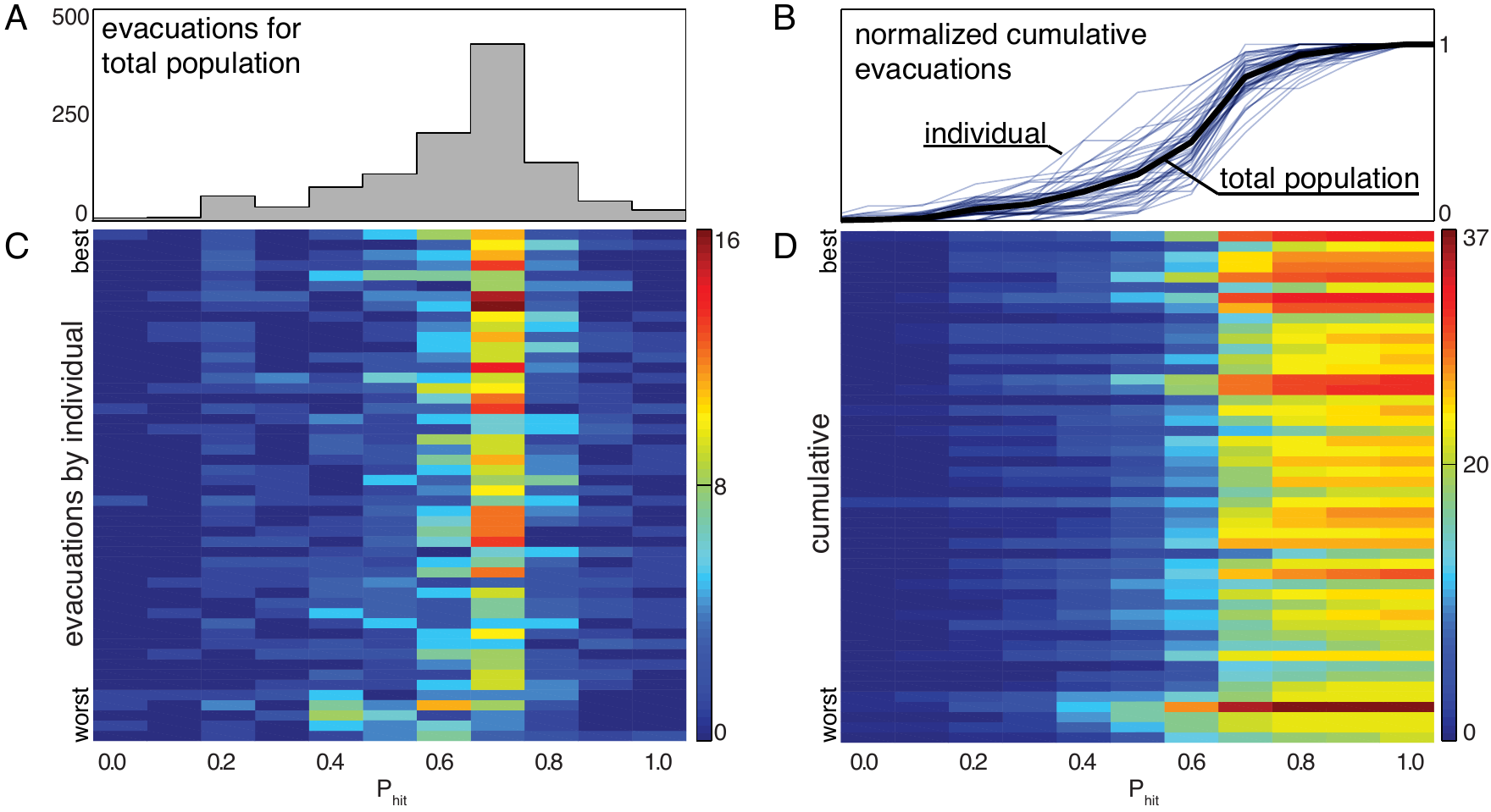}
\end{center}
\caption{{\bf The distributions of evacuations as a function of $P_\mathrm{hit}$.} Frame \textbf{A} shows the numbers of evacuations at each of the eleven values of $P_\mathrm{hit}$. The distribution is peaked at $P_\mathrm{hit}=0.7$. Frame \textbf{B} presents the normalized cumulative evacuation curves with individuals shown in blue and the population as a whole (the running sum of the distribution in \textbf{A}) in black. This provides a summary of the heterogeneity in evacuation decisions. Frame \textbf{C} shows the evacuations for each individual participant. Here we illustrate results for the highest scoring participant at the top and the lowest scoring participant at the bottom. We see a trend that the higher scoring participants evacuated more consistently at $P_\mathrm{hit}=0.7$, and the lowest scoring individuals have greater spread in the $P_\mathrm{hit}$ values at which they evacuated. Frame \textbf{D} gives the cumulative evacuations, a running sum of the data presented in \textbf{C}. We see that higher scoring individuals evacuate more readily, with the noted exception of the fourth worst scoring participant, who tended to evacuate much earlier than the others; a strategy that resulted in many unsuccessful evacuations.}
	\label{players1011}
\end{figure*}

\paragraph{Evacuation decision tied to disaster likelihood.}
Disaster likelihood values strongly influenced decision making, as shown in
Fig.~\ref{players1011}\textbf{A}. Here we see each observed evacuation
grouped by $P_\mathrm{hit}$ value at the time of evacuation. The distribution
has a sharp peak at $P_\mathrm{hit}=0.7$. The cumulative distribution is shown in Figure
\ref{players1011}\textbf{B} (black) and
indicates that across all scenarios, about 90\% of evacuations occurred
before $P_\mathrm{hit}$ exceeded 80\%.

\paragraph{High scoring individuals evacuate frequently.}
We observed a significant correlation between score and number of evacuations
at $P_\mathrm{hit}=0.7$ (Pearson correlation: $r=0.59$, $p=5.8 \times
10^{-6}$). The lowest scoring individuals (see
Fig.~\ref{players1011}\textbf{C}, bottom) evacuate earlier and have a greater
variation in the $P_\mathrm{hit}$ values at which they evacuate. In
Fig.~\ref{players1011}\textbf{D} we present the cumulative number of
evacuations, a running sum of the the data in
Fig.~\ref{players1011}\textbf{C}. Here we observe a relationship between the
total number of evacuations and score: highest scoring participants (top) are
more likely to have a higher number of total evacuations than lower scoring
participants (bottom). We confirmed this observation by calculating the
Pearson correlation between score and total number of evacuations: $r=0.39$
with $p=0.005$. A notable exception to this trend is the fourth lowest
scoring participant who also has the highest number of evacuations.
Interestingly, this participant tended to evacuate much earlier than the
other participants, resulting in many erroneous evacuations and therefore a
lower cumulative score.

\section*{Analysis}
Following the experimental observations described above, our objective
is to identify a model for evacuation decision making that can be used to
quantitatively capture the main features of population level behavior (this
section) and the heterogeneity of individual behavior (next section). The model will allow us to infer how the different experimental variables affect evacuation decision making. Our
strategy uses data from the behavioral experiment to determine a decision
model that depends on a few key state variables in the experiment (e.g., the
probability of the disaster event $P_\mathrm{hit}$). Based on summary
statistics of evacuation behavior, we identify the functional form of the
model and quantitatively estimate parameters. We then evaluate the accuracy
of the model for predicting evacuations using state variables and detailed
time trajectories from each individual run of the experiment. Our approach
enables a concrete validation of our model, and provides direction for future
experiments and large scale simulations of population behavior in similar
scenarios.

Determining the dynamics of decision making strategies from the distribution
of evacuations (Fig.~\ref{players1011}\textbf{A}) is a complex problem that
can be confounded by various factors including the distribution of
$P_\mathrm{hit}$ values observed by a participant and individual differences
in reaction time. To account for these factors we introduce a rate model
relating the number of participants evacuated to the number of participants
AtHome, and determine how state variables such as $P_\mathrm{hit}$ affect the
rate.

As $P_\mathrm{hit}$ changes every second in our scenarios, it is natural for
us to examine the data in one second intervals, within which $P_\mathrm{hit}$
is constant.  We then define two indicator functions that enable us to
quantify the number of participants evacuated and the number of participants
AtHome.
First, we define the indicator variable $h_{l,i}^{(r)}=1$ if participant $l$
was AtHome at the start of the interval $i$ during run $r$, and
$h_{l,i}^{(r)}=0$ otherwise (i.e., the participant had already evacuated).
Second, we define the indicator variable $j_{l,i}^{(r)}=1$ if participant $l$
evacuated during interval $i$ on run $r$, and $j_{l,i}^{(r)}=0$ otherwise.
These quantities are related by the equation:
\begin{equation}
	j_{l,i}^{(r)}=h_{l,i}^{(r)}-h_{l,i+1}^{(r)}.
\end{equation}

We approximate an individual's decision to evacuate as a Bernoulli process in
the following way. First we note that when $h_{l,i}^{(r)}=1$, we can model
the probability of evacuating during the interval $i$ as a \textit{rate},
denoted $\theta_{l,i}^{(r)}$, where $\theta_{l,i}^{(r)}\in [0,1]$. We treat
the observed value for the indicator variable $j_{l,i}^{(r)}$ as one sample
of an underlying stochastic process that can take a value of either $0$ or
$1$. A single sample of the data provides a poor estimate of the rate
$\theta_{l,i}^{(r)}$. However, by modeling the data as a Bernoulli process,
we can estimate the variance in rate, based on our limited number
of observations. This approach enables us to derive a decision model without
overestimating our confidence in small samples of data.

We hypothesize that $\theta_{l,i}^{(r)}$ varies in a predictable manner
according to a small set of state variables that capture the essential
decision parameters in the experiment. To uncover these trends, we partition
the data in a number of ways in this and the following section. In this
section, we combine data for all the participants to obtain aggregate rates for
the population as a whole, and in the following section, we consider
heterogeneity in the evacuation rates of individual participants.

We begin by aggregating the data for specific disaster likelihoods
$P_\mathrm{hit}$, which in the experiment can take on values $\nu \in \{0.0,
0.1, 0.2, \dots, 0.9, 1.0\}$. For each possible value $\nu$, we determine the
total number of intervals in the aggregate experiment where a participant who
is AtHome observed $P_\mathrm{hit}=\nu$:
\begin{equation}
	H_\nu = \sum_l \sum_r \sum_{i:P_\mathrm{hit}=\nu} h_{l,i}^{(r)}.
\end{equation}
We likewise determine the total number of times such participants then
evacuated:
\begin{equation}
	J_\nu = \sum_l \sum_r \sum_{i:P_\mathrm{hit}=\nu} j_{l,i}^{(r)}.
\end{equation}

We use the uppercase $\Theta_\nu$ to indicate the evacuation rate for each
value $P_\mathrm{hit}=\nu$. If we think of $J_\nu$ as a random
variable (modeled as a sum of Bernoulli variables) given $\Theta_\nu$ and
$H_\nu$, then $J_{\nu}$ has a binomial distribution. Conversely, the
likelihood of $\Theta_\nu$ given $H_\nu$ and $J_\nu$, has a
Beta$(\alpha,\beta)$ distribution \cite{otto_biologists_2007}, with parameters $\alpha = J_\nu + 1$ and
$\beta = H_\nu - J_\nu + 1$. We thus \textit{measure} rates from the data
using the expected value of this Beta distribution:
\begin{equation}
	\Theta_\nu = \mathbb{E} \ \mbox{Beta}(J_\nu + 1,H_\nu - J_\nu + 1) = \frac{J_\nu+1}{H_\nu+2}. \label{eq:A}
\end{equation}
The standard deviation of these estimates is given by:
\begin{equation}
	\sigma(\Theta_\nu) = \sqrt{\frac{(J_\nu+1)(H_\nu-J_\nu+1)}{(H_\nu+2)^2(H_\nu+3)}}.\label{eq:variance}
\end{equation}
Given an abundance of data, the measured rate converges to the more intuitive
fraction of evacuations $J_\nu/H_\nu$. However, when data is limited the
approach described above yields a more accurate description of the evacuation
behavior.

\begin{figure}[htbp]
\begin{center}
	\includegraphics[scale=0.90]{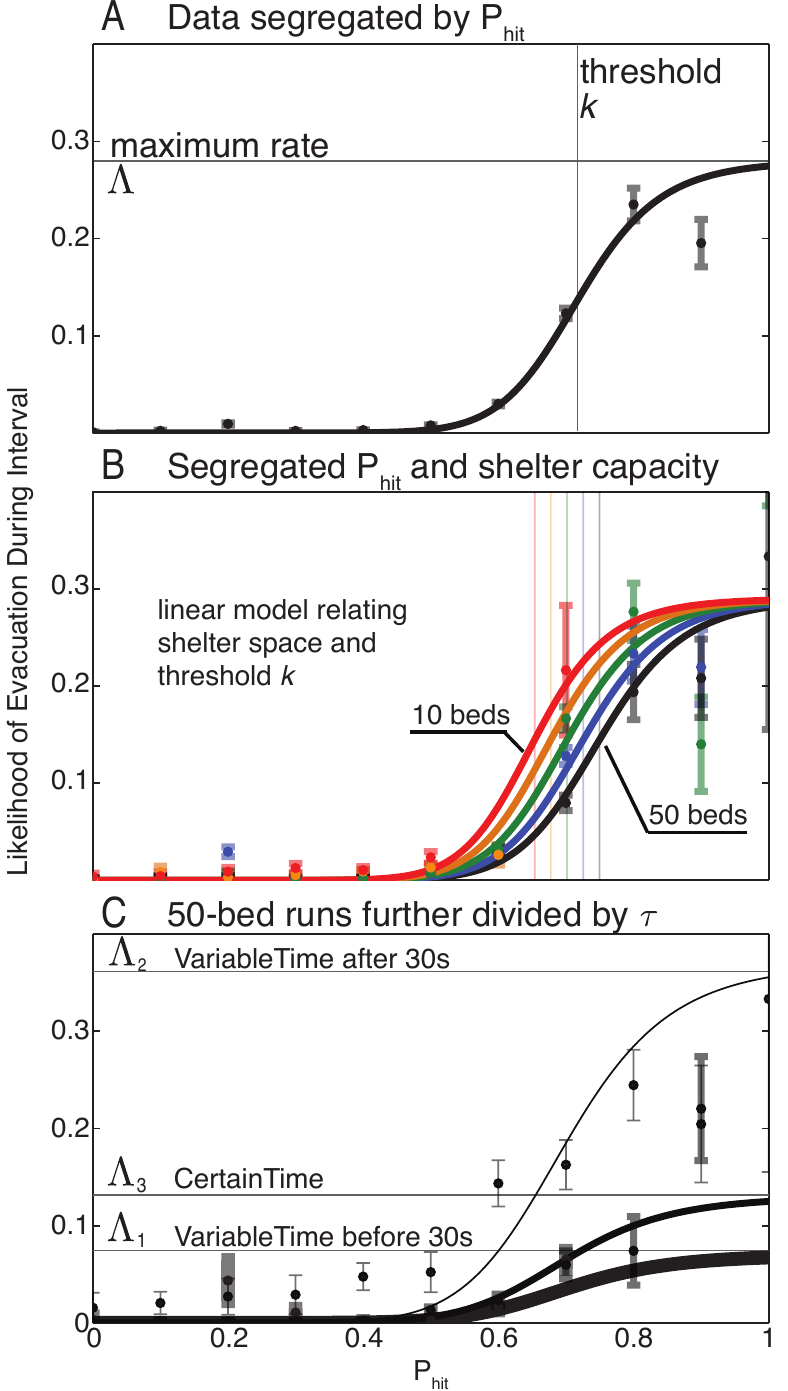}
\end{center}
\caption{{\bf Model rate laws and their variation with shelter capacity and time pressure.} In
\textbf{A} we plot the measured rates for data partitioned only by
$P_\mathrm{hit}$ (black dots with grey bars for standard deviation), along
with the best fit model (Eq.~\ref{eq:Hill}). In \textbf{B} we plot the measured
rates for the data further partitioned by shelter capacity $s$, along with the
best fit model where the mean threshold $k$ is a linear function of $s$
($k=ms+b$). Line color indicates shelter capacity: $s=10$ (red; top),
$s=20$ (orange), $s=30$ (green), $s=40$ (blue), and $s=50$ (black; bottom).
Not all $P_\mathrm{hit}$ values were observed in all $s$
value scenarios. As bed number decreases, the rate curve shifts left, giving an
increase in evacuation rate at the same $P_\mathrm{hit}$. The model in \textbf{B} displayed systematic inaccuracies requiring partitioning the data into three different time scenarios ($\tau=1$ before 30 seconds in 30 second or greater runs, $\tau = 2$ after 30 seconds in those runs, and $\tau = 3$ for 60 second runs). In \textbf{C} we plot only the 50-bed curves for the three scenarios and note that the rates for $\tau = 3$ lie between $\tau=1$ and 2.\label{fig:Hill}}
\end{figure}

Fig.~\ref{fig:Hill}\textbf{A} shows the estimated $\Theta_\nu$ rates (black
dots) associated with the 11 possible values $\nu$ of the disaster likelihood
$P_\mathrm{hit}$. We observe that the rates increase approximately
monotonically with $P_\mathrm{hit}$ in a manner that is reminiscent of a Hill
function \cite{shuler_bioprocess_2001}. We therefore model $\Theta_\nu$ using
the following functional form:
\begin{equation}
	\mu(P_\mathrm{hit}) = \Lambda \frac{P_\mathrm{hit}^n}{P_\mathrm{hit}^n + k^n},\label{eq:Hill}
\end{equation}
which enables us to describe the decision making dynamics of the population
using three parameters. First, $\Lambda$ denotes the maximum evacuation rate;
when $P_\mathrm{hit}$ is large, $\mu$ saturates to this value. $\Lambda$ can
therefore be used to estimate how quickly participants are able to react to
rapidly changing conditions. Second, the threshold parameter $k$ represents
the half maximum value of $P_\mathrm{hit}$, i.e., $\mu(k) = \Lambda / 2$.
Third, the Hill-parameter $n$ dictates the steepness of $\mu$ at $k$. For
large values of $n$ (e.g., $n>20$), $\mu(P_\mathrm{hit})$ is threshold-like,
being approximately $0$ for $P_\mathrm{hit}<k$, and approximately $\Lambda$
for $P_\mathrm{hit}>k$. For smaller values of $n$ the transition is more
gradual. Threshold policies have been extensively studied in previous work
and are postulated to accurately characterize individual decision making
behaviors in a variety of scenarios
\cite{Granovetter1978,macy_chains_1991,Watts2002,BassettEtAl2012-PRE}.

All models used in the manuscript are fit to the data by evaluating the
measured rates at each value $\nu$ of the disaster likelihood to obtain
$\mu_\nu$. We then vary $\Lambda, k$, and $n$ to maximize the expression:
\begin{eqnarray}
	&\sum_{\nu}\left[(H_\nu-J_\nu)  \ln(1-\mu_\nu)+ J_\nu \ln(\mu_\nu)\right],\label{eq:score}
\end{eqnarray}
a fit directly to the $H_\nu$ and $J_\nu$ values, not the $\Theta_\nu$ values. This expression is derived through maximum likelihood estimation \cite{bevington_data_2002} for Beta distributed measurements. The more common $\chi^2$ minimization for curve fitting is similarly derived from maximum likelihood estimation for Gaussian distributed measurements \cite{bevington_data_2002}, and our formula serves the corresponding role.

Fitting our model to the measured rates in Fig.~\ref{fig:Hill}\textbf{A}, we
obtain $k=0.72\pm 0.03$, $\Lambda=0.28\pm 0.06$, and $n=11.9\pm 1.4$. The standard deviations reported here were obtained via bootstrapping \cite{press_numerical_1992} where we constructed synthetic data sets by randomly selecting 47 runs with replacement from the original data, then aggregating the data and fitting the model to the synthetic data using the method described above. The best fit model is plotted in Fig.~\ref{fig:Hill}\textbf{A} (solid black line). For most values of $P_\mathrm{hit}$, we find that this
model accurately captures the observed behavior. However, we also observe
systematic variations between the model and the experimental data. One set of
variations appears to stem from shelter capacity while the other appears to
stem from temporal urgency for the evacuation decision. 

To examine the role of shelter capacity $s$ in decision making, we aggregate
the data for each of the $11$ disaster likelihoods $P_\mathrm{hit}$ at each
of the $5$ values of shelter capacity $s$. We adapt our use of the subscript
$\nu$ to now indicate this finer-grained aggregation into $11\times5$ sets of
data. The measured rates confirm our expectation that evacuation rates were
high when shelter space was scarce and low when shelter space was abundant
(see Fig.~\ref{fig:Hill}\textbf{B}).

To model the role of shelter capacity in modulating the average form of the
evacuation decision, we consider two families of Hill functions based on our previous fits: one family drawn from variations in $\Lambda$ and a second family drawn from variations in $k$. To guide our choice between these two alternatives, we consider optimal decision making behavior. If shelter space is abundant and information is precise, the optimal evacuation decision rule will be a threshold-like function in which the value of the threshold is just below $P_\mathrm{hit}=1.0$. This behavior ensures that the individual evacuates when there is near certainty that the disaster will hit the community. If instead there is very limited shelter space and the costs of the two possible incorrect decisions are equal, the expected evacuation decision rule will also be a threshold-like function, but in this case the value of the threshold will be just above $P_\mathrm{hit}=0.5$. This behavior ensures the best chance of getting a bed in the shelter, which is the lowest loss associated with a wrong decision.

Because the threshold value appears critical for optimal decision making
behavior in scenarios of both abundant and scarce shelter space, we choose
the family of Hill functions obtained from varying $k$. We find that the
following linear model of $k$ versus $s$:
\begin{equation}
	\mu(P_\mathrm{hit},s) = \Lambda \frac{P_\mathrm{hit}^n}{P_\mathrm{hit}^n + (ms+b)^n},\label{eq:Hill2}
\end{equation}
fits the data well. In Fig.~\ref{fig:Hill}\textbf{B}, we show the set of
curves extracted for the best fit to the model in \eqref{eq:Hill2} alongside
the raw empirical data. The best fit values for $k = ms+b$ are $m=0.0024$ and
$b=0.28$.

To test the accuracy of this model and to identify systematic differences
between the best fit model and the data, we compared the predictions of this 
model to the data, and found a systematic trend whereby we overestimate the
number of evacuations occurring prior to 30 seconds in VariableTime runs and underestimated the number of evacuations occurring after 30 seconds in those runs. The difference between actual 
and predicted evacuations was profound and the shift between overestimating to underestimating 
was abrupt, shifting at exactly the 30 second mark in nearly every VariableTime run. 
These results show that an individual's behavior is additionally influenced by temporal
urgency.

To quantify the effect of temporal urgency, we extend our model in the
following way. As in the previous versions of the model, we aggregate the
data for each of the $11$ $P_\mathrm{hit}$ values at each of
the $5$ values of shelter capacity $s$. However, in this case we additionally
aggregate data for the following $3$ separate cases with differing temporal
urgency: prior to 30 seconds in VariableTime runs
($\tau=1$), after 30 seconds in those runs ($\tau=2$), and all data in
CertainTime runs ($\tau=3$). We again adapt our use of the
subscript $\nu$ to now indicate this even finer-grained aggregation into
$11\times5\times3=165$ sets of data. 

To determine if temporal urgency had a more significant effect on $\Lambda$ or on the threshold parameters ($m$, and $b$), we fit the model equation in Eq.~\ref{eq:Hill2} independently to the 3 $\tau$ cases. From these fits and the confidence intervals on the parameter estimates we were able to determine that the variation of $\Lambda$ with temporal urgency was more significant than the variation of $n$, $m$, or $b$. We therefore constrained variation with temporal urgency to $\Lambda$, adopting a six parameter model:
\begin{equation}
	\mu(P_\mathrm{hit},s,\tau) = \Lambda_\tau \frac{P_\mathrm{hit}^n}{P_\mathrm{hit}^n+(ms+b)^n},\label{eq:6param}
\end{equation}
which has three $\Lambda_\tau$ values. The best fit values are presented in Table \ref{tab:1}. 

\begin{table}[h]
\begin{center}
\begin{tabular}{| l c  l l |}
\hline
Parameter & Symbol & Value & STD \\
\hline
Hill-coefficient 			&	$n$                 & ~9.3  			&$\pm 1.3$ \\
\hline
Maximum rates: &&&\\
~~ $\tau=1$	&	$\Lambda_1$         & ~0.07  						& $\pm 0.02$  \\
~~ $\tau=2$	&	$\Lambda_2$         & ~0.37  						& $\pm 0.07$  \\
~~ $\tau=3$	&	$\Lambda_3$         & ~0.13  						& $\pm 0.04$  \\
\hline
Threshold parameters:		&($k=ms+b$)& &\\
~~Offset 						& 	$b$ 				& ~0.60  	& $\pm 0.05$ \\
~~Proportionality const.		& $m$            & ~2$\times10^{-3}$& $\pm 1\times10^{-3}$ \\
\hline
\end{tabular}
\end{center}
\vspace{-5mm}
\caption{Parameter Estimates for the the model in Eq.~\ref{eq:6param}, with standard deviations obtained via bootstrapping \cite{press_numerical_1992}.
\label{tab:1}}
\end{table}

Figure \ref{fig:Hill}\textbf{C} illustrates the measured rates and model curves for a characteristic subset of the data (runs with 50 beds) for each of the three
time windows ($\tau=1,2,3$). For this partitioning of the data both the first 30 seconds of VariableTime runs ($\tau=1$) and the full 60 seconds of CertainTime runs ($\tau=3$) are described by similar low evacuation rates $\Lambda_1=0.07$ evacuations/second and $\Lambda_3=0.13$ evacuations/second, respectively. Both of these are significantly smaller than the corresponding rate $\Lambda=0.28$ evacuations/second for original aggregated data (Figure \ref{fig:Hill}\textbf{A}) as well as the rate $\Lambda_2=0.37$ evacuations/second observed after 30 seconds in the VariableTime runs ($\tau=2$). The increase in rate during the uncertain window in the VariableTime runs reflects a high temporal urgency associated with a disaster that could strike at any moment. It also suggests participants will respond quickly to changing $P_\mathrm{hit}$ values under these conditions.

\begin{figure*}[htbp]
\begin{center}
	\includegraphics[scale=1.0]{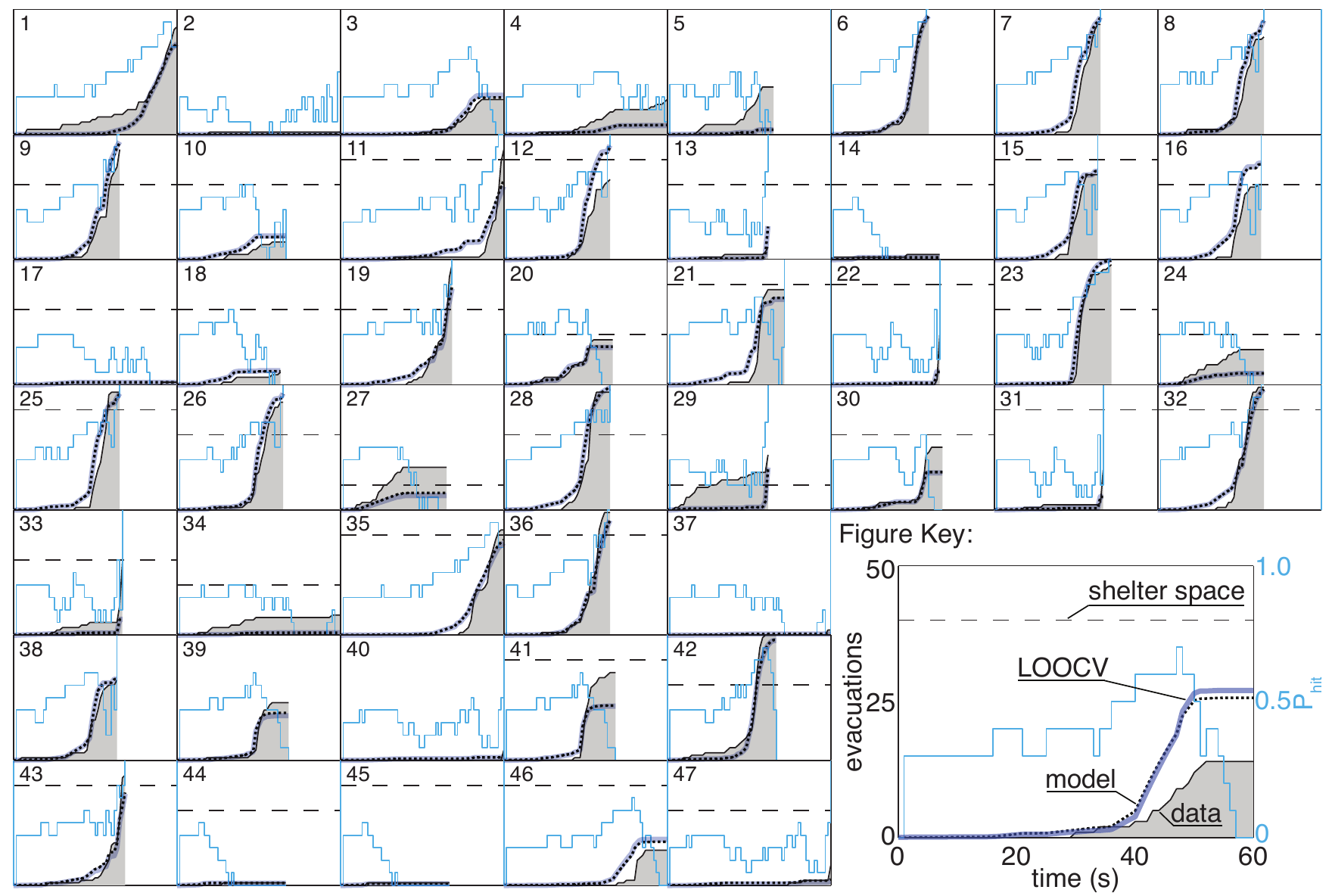}
\end{center}
	\caption{{ \bf A comparison between data and simulation for the 47 scenarios and the best fit six-parameter model} defined in Eq.~\ref{eq:6param}. At each second the $P_\mathrm{hit}$ value (blue), the shelter capacity, and the time scenario determine the rate used in the simulation, and the expected number of evacuations is calculated. The model was fit to estimated rates (Eq.~\ref{eq:A}), not to the time series data shown here. This extends the ability of the model to predict untested scenarios. To illustrate the predictive capability we also plot the {\it leave-one-out cross-validation} (LOOCV) predictions (violet curves). If the model were over-fit, the LOOCV curves would have significant deviation from the full model. The reduction from 2820 rates in the data to a six-parameter model generated a model with surprising accuracy. The following runs had identical $P_\mathrm{hit}$ trajectories: (1,35), (3,46), (8,25), (9,36), (12,26), (13,29), (14,44,45), (15,16,38), (19,43), (22,31,33), (34,37), (39,41), (40,47).}
	\label{fig:allRuns}
\end{figure*}

The relatively low values of  $\Lambda_1$ and $\Lambda_3$ are likely due to the fact that in these cases the disaster strike is only possible in the last time increment of these partitions, a low temporal urgency. In each case, urgency increases towards the end of the interval, and this occurs to a greater degree for $\tau=3$  (CertainTime) than for $\tau=1$ (first time window in VariableTime). In CertainTime runs, the scenerio terminates at exactly 60 seconds, so in this case the last observed 
$P_\mathrm{hit}$ value describes the likelihood of a strike \textit{at} 60 seconds, whereas in the first 30 seconds of the VariableTime runs the value of $P_\mathrm{hit}$ at the end of the interval
reflects the probability of a Hit not necessarily in the next time increment, but rather at some time within the uncertain 30 second window. We expect this distinction underlies our observation that
$\Lambda_3>\Lambda_1$.

\subsection*{Simulations}
We test our decision model by using it to simulate evacuation behavior for
the 47 scenarios in the behavioral experiment. The appropriateness of our
model can then be quantified by the difference between simulated and observed
behavior, with small differences indicating that our model could be used as a
generative model in future numerical studies.

In the experiment, each scenario is characterized by a shelter capacity $s$
and time pressure $\tau$, as well as a prescribed sequence of disaster
likelihood values $P_\mathrm{hit}$. Using our decision rule, we can compute
the expected rate of evacuations at each instantaneous value of
$(s,~\tau,P_\mathrm{hit})$. If we initialize every simulation with 50
individuals at home ($H_{0}^{(r)}=50$), we can compute the expected
number of people AtHome in each interval ${\langle H_i^{(r)} \rangle}$ using:
\begin{equation}
	\left<H_{i+1}^{(r)}\right> = \left[1-\Lambda_\tau \frac{P_\mathrm{hit}^n}{P_\mathrm{hit}^n+(ms+b)^n}\right] \left<H_{i}^{(r)}\right>.
\end{equation}
In the paragraphs below, we comment briefly on several key results from our
simulations (see Fig.~\ref{fig:allRuns}).

\paragraph{Decision model accurately describes experimental observations.} In the majority of scenarios the simulated behavior has very little deviation from the observed behavior. This result is striking because our  model aggregates the data over all participants over all scenarios to a reduced set of six parameters, with no time resolution aside from separation into the three bins associated with the different time pressure variables. In the majority of scenarios the simulated evacuation behavior is qualitatively, and in many
cases quantitatively, matched to the observed behavior of experiment
participants.

As a check that we have not over-fit the model, we have performed a {\it leave-one-out cross-validation} (LOOCV) \cite{hastie_elements_2009}, where for each of the 47 runs, we exclude the data from that run, and see how the model trained on the other 46 runs predicts the outcome. The LOOCV results (Fig.~\ref{fig:allRuns}, violet curves) were nearly identical to the predictions of the full model (Fig.~\ref{fig:allRuns}, dotted curves), indicating that the model is not over-fit. This result also suggests that the model will predict the outcome of other scenarios with the same accuracy of the simulations shown here, assuming that the $P_{hit}$ trajectories are created using the same rules.

We begin our description of Fig.~\ref{fig:allRuns} with the three runs where participants had the most 
success, 36, 44, and 45. As can be seen here and in Fig.~\ref{playerpayoffs}\textbf{A} (far right), all but a single individual made the correct evacuation decision in these runs. In run 36, the disaster had a very predictable trajectory, gradually increasing in $P_\mathrm{hit}$ before eventually striking. In runs 44 and 45, the disaster had a poor likelihood of
striking and $P_\mathrm{hit}$ decayed fairly rapidly. In contrast, the most difficult run was number 42.  The $P_\mathrm{hit}$ trajectory in this run peaked at 0.9 before the chance of a disaster strike rapidly decayed and the run ended with a Miss. As can be seen here and in Fig.~\ref{playerpayoffs}\textbf{A} (far left) every participant was left either InShelter or InTransit.

\paragraph{We observed sub-optimal decision making.}
In general, the optimal decision to evacuate in a given scenario depends not only on the likelihood and volatility of the underlying disaster process, as well as on the loss matrix, but also on the shelter capacity and the decisions of other individuals. However, scenarios 1, 2, 3, 4, 37, and 40 are unusually simple in that participants knew that these scenarios would each last exactly 60 seconds, and that there was adequate shelter capacity for all participants. These two simplifying factors ensured that the actions of other participants had no direct effect (though they could presumably influence behavior, e.g.~peer pressure). In these scenarios, it would be optimal to wait until immediately before the potential disaster strike to evacuate.
As Fig.~\ref{fig:allRuns} indicates, in scenarios 1, 3, and 4, participants
did not follow the optimal strategy; rather a significant number of
participants evacuated well before the end of the scenario.  In fact, many
participants evacuated after only approximately 30 seconds. This behavior
proved costly for them in scenarios 3 and 4. Scenarios 2, 37, and 40 are less
conclusive because the strike likelihood $P_\mathrm{ hit}$ in these scenarios
never exceeded 0.5 (and the disaster did not hit), making it relatively easy
to decide not to evacuate.

\paragraph{Participant behavior adapts over time.}
By construction, several scenarios contained identical $P_\mathrm{hit}$
trajectories but differed in other parameters. Among these ``repeated''
disasters, we observe evidence of learning with regard to time pressure. In
runs 1, 3, and 8 there were some unnecessarily early evacuations, but
participants waited longer to evacuate in the corresponding runs occurring
later in the experiment (runs 35, 46 and 25).

This observed adaptation could be explained either by effects of time
pressure or by effects of strike likelihood. To determine the dominant driver
of the adaptation, we compared the evacuation rates in runs 1-8 with those in
runs 37-40 to determine whether there was evidence for adaptation in
decision making strategies. 
While these runs differed in strike likelihood, the measured rates observed in the two groups
did not show a significant change at high
$P_\mathrm{hit}$ values. This suggests that although participants seemed to
adapt their strategies in relation to time pressure, they did not adjust
their behavior in relation to strike likelihood.

\paragraph{Unexpectedly extreme sensitivity to shelter capacity.}
In each of scenario runs 27 and 29, shelter beds were scarce (10 beds for 50
people) and more participants evacuated early in the scenario than our model
predicted. It is possible that either (1) our linear model of the variation of the
threshold $k$ with shelter capacity $s$ is inadequate when shelter space is very
scarce, (2) that time pressure affects player behavior before 30 seconds in VariableTime runs with low shelter capacity, or (3) the participants were reacting to each of these scenarios also immediately following runs in
which a large number of individuals evacuated after the shelter was full,
leaving those individuals stuck InTransit (runs 26 and 28). The early
evacuations in runs 27 and 29 could therefore be a reaction to participants
being caught InTransit in the previous run. We are unable to discriminate between these three possibilities with this data set; we leave this for future work.

\subsection*{Individual Variation} \label{sec:individual}
Our success in identifying a decision making model that captures the observed
collective evacuation behavior in the experiment led us to test whether a
similar method could differentiate between individual decision making
strategies. In the previous analyses, we combined data for all of the
participants, which enabled us to fit the model to several experimental
variables. Because the evacuation data for individual participants is
relatively sparse, here we focus exclusively on the influence of the disaster
likelihood $P_\mathrm{hit}$ in decision making and do not separately consider
the effect of shelter capacity or time pressure.

\begin{figure*}[htbp]
\begin{center}
	\includegraphics[scale=1]{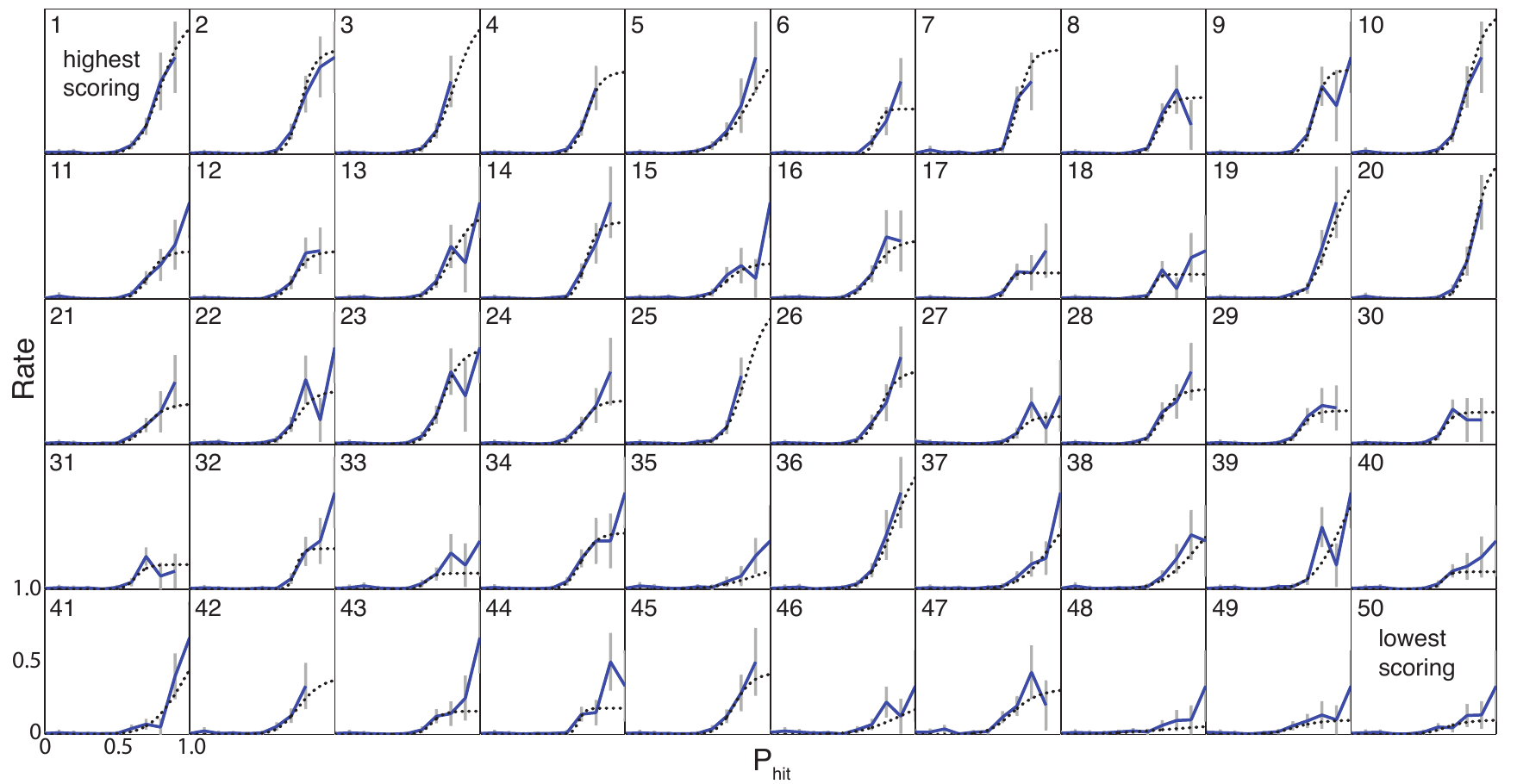}
\end{center}
	\caption{{\bf A comparison between the decision making model and data from the
behavioral experiment} for each participant, ranked according to cumulative
score. Evacuation rates for each individual at each $P_\mathrm{hit}$ value
were measured using Eq.~\ref{eq:A}. These values are plotted in blue
accompanied by the estimated standard deviations for each point (grey bars)
calculated based on Eq.~\ref{eq:variance}. Hill functions were fit for each
individual using the routine described in Eq.~\ref{eq:score} (dotted black).
Higher evacuation rates tend to result in higher scores. The fits give a
significant correlation between evacuation rate $\Lambda$ and score (Pearson
$r =0.41$, $p=0.0028$). Moreover, individuals who evidencing higher financial
risk attitude scores (i.e., more risk seeking) have higher thresholds for
evacuation $k$ than individuals evidencing lower financial risk attitude
scores (Pearson $r=0.30$, $p=0.03$).}
	\label{fig:individuals}
\end{figure*}

To extend the collective decision making model to individuals
we estimated the evacuation rates for each participant at each
$P_\mathrm{hit}$ value using Eq.~\ref{eq:A}. We show this data in
Fig.~\ref{fig:individuals}, where individuals are ranked by score from
highest scoring (top left) to lowest (bottom right). Some individuals had
as few as 9 measured rates, as they consistently evacuated before
$P_\mathrm{hit}\geq 0.9$ (see truncated curves in
Fig.~\ref{fig:individuals}).

Comparing the raw data in Fig.~\ref{fig:individuals} for individuals with the
corresponding measured rates for the aggregate population shown in
Fig.~\ref{fig:Hill} illustrates an interesting deviation in the measurements
at high values of $P_\mathrm{hit}$. For the aggregate population there is a
significant and somewhat counterintuitive drop in measured rate from
$P_\mathrm{hit}=0.8$ to 0.9; the value of the measured rate represented by
the data points at $P_\mathrm{hit}=0.9$ lies below the value represented at
$P_\mathrm{hit}=0.8$. However, while non-monotonicity is observed on the scale 
of individuals the trend is not systematic
(see Fig.~\ref{fig:individuals}). The difference
between the population and individual fits suggests that the observed drop in
the measured rate at high $P_\mathrm{hit}$ in aggregate data is driven by
heterogeneity in the population. Participants with high evacuation rates tend
to leave before $P_\mathrm{hit}\ge 0.9$. Those who remain and observe high
values of $P_\mathrm{hit}$ typically display low evacuation rates, thereby
biasing the summary rates measured at the population scale.

To capture individual decision making strategies, we fit a three-parameter
Hill function (Eq.~\ref{eq:Hill}) to each individual's measured rates using
Eq.~\ref{eq:score}. As shown in Fig.~\ref{fig:individuals}, the best fit
models based on the Hill function capture the measured rate curves of each
participant with striking accuracy.

\paragraph{Higher evacuation rates accompany better performance.}
The wide range of participant decision making behavior is clearly visible in
Fig.~\ref{fig:individuals}. The variability is especially apparent when we
compare the highest scoring individuals with the lowest scoring individuals.
The highest scoring participants exhibit rates that increase sharply and
monotonically, approximately beginning at $P_\mathrm{hit}=0.7$. The lowest
scoring individuals rarely evacuate; we observe flat evacuation rate curves, with measured rates that are relatively much lower and less systematic in their variations compared to high scoring individuals. As is apparent from the accuracy of the fits, this distinction is well captured by our model.

\begin{figure*}[htbp]
\begin{center}
	\includegraphics[scale=1]{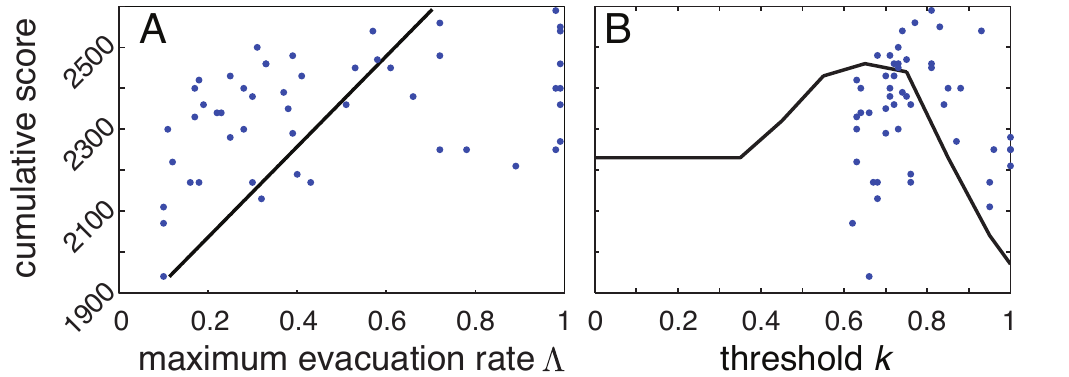}
\end{center}
\caption{{\bf Best fit models provided values for the maximum
evacuation rate $\Lambda$ and threshold parameter $k$ for each individual.} \textbf{A} The distribution of $\Lambda$ values across
participants spanned almost the full range from $0$ to $1$.  Here we observe
a significant correlation between $\Lambda$ values and cumulative score
across participants (Pearson $r =0.41$, $p=0.0028$). This result provides
statistical support for the apparent tendency for high scoring individuals to
also display higher rate values (see Fig.~\ref{fig:individuals}). \textbf{B} cumulative score vs threshold parameter (blue dots) had no significant linear correlation. A strict thresholding strategy (black curve), where a model player would immediately evacuate once $P_{hit}$ exceeded their threshold, helps to explain the lack of linear correlation. If a threshold is set too high, it results in many AtHome Hits while too low results in InShelter Misses. There is a maximum cumulative payment for strict thresholding between 0.6 and 0.7. We see that participants typically had thresholds above this range and scored higher than the expected payoff (blue dots). This is likely a result of participants incorporating time pressure and scarcity into their decisions, having reductions in score from a low $\Lambda$, and variability in having a non-threshold (low $n$) strategy.}
	\label{fig:ScoreAndRate}
\end{figure*}

A fundamental goal of our experiment was to identify psychological and
behavioral predictors of individual performance. First, we ask whether parameter values from the best fit models on individual participants could be related to behavioral performance in the experiment. The best fit models yielded rates $\Lambda \in [0,1]$, with values for every individual displayed in Fig.~\ref{fig:ScoreAndRate} \textbf{A}. Overall, we observe a significant positive correlation between the maximum evacuation rate $\Lambda$ in the best fit models and cumulative score (Pearson $r =0.41$, $p=0.0028$; see Fig.~\ref{fig:ScoreAndRate}\textbf{A}). We speculate that the maximum evacuation rate could be related to a participant's fundamental reaction time. If true, our results suggest that participants who can react quickly to rapidly changing conditions in their environment are more successful in the experiment.

As expected, we do not see a significant linear correlation between cumulative score and threshold parameter $k$. This results from a mid-range value of $k$ having an optimal effect, with low thresholds resulting in erroneous evacuations, and high thresholds resulting in disaster strikes while AtHome. To illustrate this optimum we plot the cumulative score varying $k$ for a strict threshold model (i.e.~ high $n$, $\Lambda=1$) in Fig.~\ref{fig:ScoreAndRate} \textbf{B} (black curve). Here we see that the maximum cumulative score for this type of decision model is at $0.6<k<0.7$. This calculation does not take into account shelter space or time pressure, which individuals (blue dots) used in order to get improved scores. The population as a whole had a higher threshold parameter ($k=0.72\pm 0.03$) reflecting the use of this additional information in obtaining higher cumulative scores. Decisions also had a considerable stochastic component for low $n$ and $\Lambda \neq 1$, giving more variability in scores.

\paragraph{Investment risk seeking may give higher evacuation thresholds.}
To identify potential psychological predictors of individual performance, we
tested whether risk scores in the financial domain and health \& safety domain were
related to individual differences in decision making strategies. We found a
significant relationship between $k$ and risk score in the investment domain
($r=0.30$, $p=0.03$), indicating that individuals with higher decision
thresholds tend to have more risk seeking attitudes. We interpret this result
with caution due to the possibility of Type II errors in the large number of
tests performed (3 risk scores and 3 best fit model parameters = 9 tested
correlations). However, a correlation between these two variables is
plausible; it suggests that participants who tolerate more financial risk are
more likely to wait until the disaster is imminent before evacuating.

\paragraph{Similar decision models can produce different scores.}
It is noteworthy that some low and intermediate scoring participants display reduced (binned) decision statistics, and consequently decision model parameters, that are almost identical  to those of the highest scoring participants. For example, participants 1 and 36 have very similar decision models but very different scores (2590 and 2270). This result indicates that in some cases similar decision making strategies can produce very different
performance outcomes.

Our decision model reduces the data to a single scenario parameter ($P_\mathrm{hit}$) and therefore fails to capture the other features that are likely to be important in distinguishing between individuals such as timing of the decision. Our data on the population scale suggested that time pressure and shelter capacity are important variables and likely have similar importance on the scale of  individuals. By comparing the detailed time evolution of individual runs, we observe instances where higher scoring participants tended to wait longer before evacuating than lower scoring participants,  a more successful strategy. 

While we are unable to quantify with significance these effects in the current experiment due to limited data, our model provides a tool for estimating the quantity of data needed to robustly quantify these parameters in driving individual decision dynamics.

\section*{Discussion}
The behavioral network science experiment reported in this paper quantifies
several key factors influencing individual evacuation decision making in a
controlled laboratory setting. The experiment includes tensions between
broadcast and peer-to-peer information, and contrasts the effects of temporal
urgency associated with the imminence of the disaster and the effects of
limited shelter capacity for evacuees. In this section we summarize
our key findings, discuss several methodological considerations, and describe
implications for future work.

\subsection*{Predictive, scalable model of collective and individual
human decision making}

Based on empirical measurements of the cumulative rate of evacuations as a
function of the instantaneous disaster likelihood, we developed a
quantitative model for decision making that captures remarkably well the main features of
observed collective behavior across the 47 disaster scenarios.
Moreover, we are able to capture the sensitivity of individual and
population level decision behaviors to external pressure on resources
(limited shelter capacity) and time (imminence of disaster). Systematic
deviations from the model provide meaningful estimates of variability in the
collective response. Our analysis uncovers a temporal evolution in individual
behavior over the course of the experiment, indicative of increasing
attention and swiftness of response, and consistent with the expectation that
individuals learn from previous incidents.

Data from the experiment reveal significant heterogeneity in individual
decision making patterns captured by significant variation in model parameter
fits to participants. The results distinguish between high scoring
individuals whose decisions to evacuate are strongly linked to a tight range
of disaster likelihoods, versus others who exhibit significantly more
variable decision making patterns and did not score as well in the
experiment. Both the individuals' overall success rate in the experiment and
the decision making variables that model their behavior are correlated with
heterogeneities in individual risk attitudes, as measured by established
psychological tests.

These results suggest new directions for numerical modeling. For example,
simulation studies that extrapolate decision making strategies identified in
small groups to larger collectives could more accurately predict behavior in
large scale populations and coalitions. Additionally, simple mathematical
models are needed to better understand the tensions and tradeoffs identified
in this experiment. Effects of competing broadcast and social information in
collective decision dynamics have been investigated previously in a numerical
simulation, where individuals were represented by nodes in a network, and
obtained information from a broadcast source as well as neighboring sites in
the network \cite{BassettEtAl2012-PRE}. In that case, decision making was
modeled as a threshold on an individual state variable representing opinion,
and the opinion of each individual was updated based on a stochastic contact
rule with the broadcast source (essentially a warning that the disaster was
coming) and other individuals (who might or might not have received any
information about the disaster). The results presented in this paper suggest
important extensions to that model that (1) incorporate different types of
information from broadcast and social sources, including an underlying
physical process involving likelihood and urgency and (2) directly implement
the individual decision model developed in this study rather than assuming
the more simplistic update rule employed previously. Our current research is focused on the design of experiments that will better characterize the role of social information and network structure. 

\subsection*{Methodological considerations}

While no laboratory experiment can fully capture the tensions associated with
a true disaster, known factors influencing human risk perception and urgency
were accounted for wherever possible in the experimental design. These
include both linguistic and visual elements, which are well studied in the
psychology and risk literature. Examples include the use and representation
of disaster likelihood rather than probability, as well as scores for each
scenario represented in terms of a potential loss rather than a payoff for a
scenario.  Previous studies have shown that humans respond differently to
losses than gains \cite{Gigerenzer2002,Chib2012}, and are significantly more
accurate in decision making based on data presented as likelihoods than on
data presented as probabilities \cite{Edwards2012,BD12}.

The changing likelihood presented to the participants in this study
represents the uncertain, and highly variable physical processes that govern
the real time approach of natural disasters, such as wildfires or hurricanes
\cite{LPP06,McCaffrey2004,CS02,R08,H11,NOAA.GOV}, and that ultimately result
in either a ``Hit'' or a ``Miss'' for individual homeowners or communities.
The existence of an underlying, quantifiable process for the disaster
introduces objective parameters that govern volatility, difficulty, and
uncertainty that can be varied in the experiment. Higher volatility, as well
as variable time steps, leads to an outcome that is more difficult to
predict. Based on the rules of the process, it is possible to calculate the
likelihood of the disaster at each time increment (which is the only aspect
of the process presented to the participants in this experiment, and it is
presented at limited resolution), as well as the optimal evacuation decision
(in the absence of shelter capacity limitations) \cite{CrewsThesis}.

The details of this process were deliberately hidden from the participants,
who were only presented with the current estimated likelihood of the disaster
hitting their community, updated at one second intervals.  Our decision to
obscure most of the details from the participants was based on observations
of realistic disaster event scenarios where the public has access to limited
information about the disaster likelihood. The complexities of geophysical
events are commonly reduced to highly simplified trajectories and
``likelihoods'' when presented to the public whether it be the chances of
rain, or the chances of a disaster \cite{regnier2008}.

In any behavioral experiment, it is of interest to compare participants'
actual behavior to optimal behavior from a profit-maximization perspective.
In our experiment, the optimal evacuation time depends both on the volatility
of the disaster process and on the potentially confounding actions of other
participants. While the choice of an underlying stochastic process in
principle allows for the calculation of a limiting theoretical optimal
decision strategy \cite{CrewsThesis}, our results demonstrate that human
behavior departs from optimality at a more primitive level. As previously
discussed, even in the simplest cases where an optimal strategy is easily
obtained (i.e., where there is no competition for shelter space, and the time
of the possible disaster strike is known in advance), the participants still
act sub-optimally. This result highlights the critical importance of
uncovering predictive models of the suboptimal decision strategies that
humans employ in real and laboratory settings.

\subsection*{A framework for quantitative analysis and prediction of
human behavior in disasters}

In the development and assessment of policy for disaster mitigation and
response, human behavioral factors are often the least well quantified,
understood, and modeled. Plans for evacuation based on broadcast
communication and transportation alone can be rendered ineffective if humans
do not act as expected. In retrospective analysis of data from recent events
\cite{CS02,LPP06,DOT06,DH07,H11}, prediction and planning for human social
factors have been identified as the critical missing link in developing
effective strategies to insure safety of the population as a whole. As a
result, critical resources are diverted to individual crisis hot spots that
might have been avoided with a more effective plan, and in many cases lives
are ultimately lost.

These shortcomings motivate our investigations, which represent the initial
steps in development of a comprehensive, predictive framework that
incorporates human factors in policy and planning for disaster mitigation and
response. Success in this area mandates an iterative approach that combines
numerical modeling with controlled experiments and retrospective analysis of
data collected from actual disasters.  Our study uncovered multiple drivers
of individual decision making behavior from competing information sources.
The social network as a whole provided a source of information on shelter
occupancy, inducing a sense of urgency in the population, while the topology
of the network surrounding a given individual (i.e., the number of that
individual's neighbors) swayed the time spent engaging the social network.
Despite these influences, individual participants spent the majority of their
time consuming the broadcast information, and the disaster likelihood was the
primary factor influencing decision making strategies in the population as a
whole.

The observed tensions between the two sources of information are consistent
with empirical observations of human behavior in real disasters. Outside of
the laboratory setting, the likelihood of a disaster event is clearly a
dominant factor in any decision to evacuate, and individuals spend a great
deal of time gathering information from television and other media broadcast
sources, even if updates are slow. However, social media and peer-to-peer
communication networks are playing an increasingly important role in
transmission of early warnings by on-site observers who may communicate
observations informally via Twitter and Facebook \cite{Dabner2012} (e.g.,
news of a 2011 earthquake in the Washington D.C.~area propagated faster on
social networks than the seismic waves themselves
\cite{Oswald2011,Ball2011}).  Furthermore, in some cases, such as developing
countries, widespread access to broadcast networks may not be readily
available, necessitating that policy makers rely on social means to
communicate information updates. Future experiments will change how participants
access information in order to investigate these
situations, and elucidate the corresponding effects on behavior.

Additionally, in many (if not most) cases social factors underlie the
decisions of individuals who evacuate early or fail to evacuate even when the
disaster is upon them \cite{DH07,H11,Drabek1986}. For example, families with
small children tend to leave early, while caring for the elderly or
reluctance to leave pets behind are often cited as reasons for not
evacuating. These factors could be incorporated in future experiments using
an explicit payoff structure that rewards collective decisions of neighbors
in the social network. Another observed source of variation in evacuations
during disasters can be traced to heterogeneities in age, health, isolation,
and socioeconomic status within the population. These factors influence speed
and access to transportation, as well as potential losses associated with
assets at risk. Such sources of variation may be incorporated in our
framework by introducing explicit heterogeneity in the loss matrix and in the
scenarios accessible to a participant during the InTransit phase.

Finally, our work highlights the role that individuality plays in the
decisions of participants and their effect on collective behavior. The
distribution of risk tendencies in this experiment might be related to the
demographics of the cohort studied here (UCSB undergraduates), and future
studies utilizing different participant groups could be used to probe such a
relationship.  For example, it is reasonable to expect that older and
wealthier individuals (e.g., homeowners) might be more risk averse in this
domain than undergraduate students. Furthermore, participants who are
explicitly trained in risk management and/or operate within different
organizational structures (e.g., military officers) might employ different
decision making strategies, and a group of such participants might by
extension display a quantitatively different collective behavior profile.

Our combined use of a novel experimental paradigm and powerful
theoretical modeling techniques to identify and quantitatively
characterize individual differences in human decision making
strategies in social groups could form a critical bridge to key work
in the fields of social neuroscience \cite{Behrens2009} and
neuroeconomics \cite{Glimcher2004,Rangel2008}, which seek to describe
neurophysiological correlates of social and economic considerations
driving human decision making. Indeed, human neuroimaging studies
highlight the role of specific brain regions in economic choices and
variations in decision strategies \cite{Venkatraman2009,Kolling2012}. Individual
differences in these circuits could underlie behavioral decision
phenotypes in healthy and diseased clinical populations
\cite{Chang2012,Hartley2012}. Uncovering neurophysiological predictors
of decision dynamics in social groups would have far-reaching
implications for disaster preparation and response, marketing, and
homeland security.

\subsection*{Development of strategies to mitigate or manage collective
evacuation behavior}

The ultimate goal of our investigations is development and testing of robust
strategies for training and control of evacuations that account for human
behavior and network topologies. These objectives may be incorporated within
our framework across both broadcast and social channels. Broadcast
information may include specific timing for public release of information,
including likelihood updates and incentives as well as warnings and mandates
for evacuation. In the peer-to-peer communication network, strategies for
robust control and potential fragilities of collective behavior may be
investigated through insertion of trained ``leaders,'' who make optimal
decisions at different locations in the network, as well as through tracing
the propagation of deliberately injected misinformation and poor decisions.
Results obtained for these ``designed'' strategies may be compared to
emergent leadership that might arise when the ranking and decisions of other
individuals in the network is communicated through the social network, an
inherent source of feedback which has been traced to the initiation of
cascades in social decision making in a wide range of applications \cite{Watts2002}.

\section*{Acknowledgments}
The authors gratefully acknowledge Michael Kearns and Stephen Judd for
providing their behavioral network science experimental framework. The
authors would like to thank Mason Porter, Michael Platt, Ali Jadbabaie, Scott Grafton, Leeda Cosmides,
John Tooby, Gary Lewis, and Rachel Silvestrini for helpful conversations.
Jason Crews and Nada Petrovic made valuable contributions to the early stages
of this research. We thank Ann Hermundstad, Jessica Wirts, and Emily Swindle
for assistance in experimental setup and procedures.
This work was supported by the David and Lucile Packard Foundation, the Office of Naval Research MURI grants N000140810747 and 0001408WR20242, and the Institute for Collaborative Biotechnologies through contract no. W911NF-09-D-0001 from the U.S. Army Research Office.

\bibliographystyle{apsrev4-1} 
\bibliography{PRE_Evac3a}

\begin{thebibliography}{95}%
\makeatletter
\providecommand \@ifxundefined [1]{%
 \@ifx{#1\undefined}
}%
\providecommand \@ifnum [1]{%
 \ifnum #1\expandafter \@firstoftwo
 \else \expandafter \@secondoftwo
 \fi
}%
\providecommand \@ifx [1]{%
 \ifx #1\expandafter \@firstoftwo
 \else \expandafter \@secondoftwo
 \fi
}%
\providecommand \natexlab [1]{#1}%
\providecommand \enquote  [1]{``#1''}%
\providecommand \bibnamefont  [1]{#1}%
\providecommand \bibfnamefont [1]{#1}%
\providecommand \citenamefont [1]{#1}%
\providecommand \href@noop [0]{\@secondoftwo}%
\providecommand \href [0]{\begingroup \@sanitize@url \@href}%
\providecommand \@href[1]{\@@startlink{#1}\@@href}%
\providecommand \@@href[1]{\endgroup#1\@@endlink}%
\providecommand \@sanitize@url [0]{\catcode `\\12\catcode `\$12\catcode
  `\&12\catcode `\#12\catcode `\^12\catcode `\_12\catcode `\%12\relax}%
\providecommand \@@startlink[1]{}%
\providecommand \@@endlink[0]{}%
\providecommand \url  [0]{\begingroup\@sanitize@url \@url }%
\providecommand \@url [1]{\endgroup\@href {#1}{\urlprefix }}%
\providecommand \urlprefix  [0]{URL }%
\providecommand \Eprint [0]{\href }%
\providecommand \doibase [0]{http://dx.doi.org/}%
\providecommand \selectlanguage [0]{\@gobble}%
\providecommand \bibinfo  [0]{\@secondoftwo}%
\providecommand \bibfield  [0]{\@secondoftwo}%
\providecommand \translation [1]{[#1]}%
\providecommand \BibitemOpen [0]{}%
\providecommand \bibitemStop [0]{}%
\providecommand \bibitemNoStop [0]{.\EOS\space}%
\providecommand \EOS [0]{\spacefactor3000\relax}%
\providecommand \BibitemShut  [1]{\csname bibitem#1\endcsname}%
\let\auto@bib@innerbib\@empty
\bibitem [{\citenamefont {Oliver}\ \emph {et~al.}(1985)\citenamefont {Oliver},
  \citenamefont {Marwell},\ and\ \citenamefont
  {Teixeira}}]{oliver_theory_1985}%
  \BibitemOpen
  \bibfield  {author} {\bibinfo {author} {\bibfnamefont {P.}~\bibnamefont
  {Oliver}}, \bibinfo {author} {\bibfnamefont {G.}~\bibnamefont {Marwell}}, \
  and\ \bibinfo {author} {\bibfnamefont {R.}~\bibnamefont {Teixeira}},\
  }\href@noop {} {\bibfield  {journal} {\bibinfo  {journal} {American Journal
  of Sociology}\ }\textbf {\bibinfo {volume} {91}},\ \bibinfo {pages} {522}
  (\bibinfo {year} {1985})}\BibitemShut {NoStop}%
\bibitem [{\citenamefont {González-Bailón}\ \emph {et~al.}(2011)\citenamefont
  {González-Bailón}, \citenamefont {Borge-Holthoefer}, \citenamefont {Rivero},\
  and\ \citenamefont {Moreno}}]{gonzalez-bailon_dynamics_2011}%
  \BibitemOpen
  \bibfield  {author} {\bibinfo {author} {\bibfnamefont {S.}~\bibnamefont
  {González-Bailón}}, \bibinfo {author} {\bibfnamefont {J.}~\bibnamefont
  {Borge-Holthoefer}}, \bibinfo {author} {\bibfnamefont {A.}~\bibnamefont
  {Rivero}}, \ and\ \bibinfo {author} {\bibfnamefont {Y.}~\bibnamefont
  {Moreno}},\ }\href {\doibase 10.1038/srep00197} {\bibfield  {journal}
  {\bibinfo  {journal} {Scientific Reports}\ }\textbf {\bibinfo {volume} {1}}
  (\bibinfo {year} {2011}),\ 10.1038/srep00197}\BibitemShut {NoStop}%
\bibitem [{\citenamefont {Howard}\ \emph {et~al.}(2011)\citenamefont {Howard},
  \citenamefont {Agarwal},\ and\ \citenamefont {Hussain}}]{Howard2011}%
  \BibitemOpen
  \bibfield  {author} {\bibinfo {author} {\bibfnamefont {P.~N.}\ \bibnamefont
  {Howard}}, \bibinfo {author} {\bibfnamefont {S.~D.}\ \bibnamefont {Agarwal}},
  \ and\ \bibinfo {author} {\bibfnamefont {M.~M.}\ \bibnamefont {Hussain}},\
  }\href {\doibase 10.1080/10714421.2011.597254} {\bibfield  {journal}
  {\bibinfo  {journal} {The Communication Review}\ }\textbf {\bibinfo {volume}
  {14}},\ \bibinfo {pages} {216} (\bibinfo {year} {2011})}\BibitemShut
  {NoStop}%
\bibitem [{\citenamefont {Chen}(2011)}]{chen2011}%
  \BibitemOpen
  \bibfield  {author} {\bibinfo {author} {\bibfnamefont {T.~M.}\ \bibnamefont
  {Chen}},\ }\href@noop {} {\bibfield  {journal} {\bibinfo  {journal} {Network,
  IEEE}\ }\textbf {\bibinfo {volume} {25}},\ \bibinfo {pages} {2} (\bibinfo
  {year} {2011})}\BibitemShut {NoStop}%
\bibitem [{\citenamefont {Khondker}(2011)}]{Khondker2011}%
  \BibitemOpen
  \bibfield  {author} {\bibinfo {author} {\bibfnamefont {H.~H.}\ \bibnamefont
  {Khondker}},\ }\href {\doibase 10.1080/14747731.2011.621287} {\bibfield
  {journal} {\bibinfo  {journal} {Globalizations}\ }\textbf {\bibinfo {volume}
  {8}},\ \bibinfo {pages} {675} (\bibinfo {year} {2011})}\BibitemShut {NoStop}%
\bibitem [{\citenamefont {Lotan}\ \emph {et~al.}(2011)\citenamefont {Lotan},
  \citenamefont {Graeff}, \citenamefont {Ananny}, \citenamefont {Gaffney},
  \citenamefont {Pearce},\ and\ \citenamefont {Boyd}}]{LotanEtAl2011}%
  \BibitemOpen
  \bibfield  {author} {\bibinfo {author} {\bibfnamefont {G.}~\bibnamefont
  {Lotan}}, \bibinfo {author} {\bibfnamefont {E.}~\bibnamefont {Graeff}},
  \bibinfo {author} {\bibfnamefont {M.}~\bibnamefont {Ananny}}, \bibinfo
  {author} {\bibfnamefont {D.}~\bibnamefont {Gaffney}}, \bibinfo {author}
  {\bibfnamefont {I.}~\bibnamefont {Pearce}}, \ and\ \bibinfo {author}
  {\bibfnamefont {D.}~\bibnamefont {Boyd}},\ }\href@noop {} {\bibfield
  {journal} {\bibinfo  {journal} {International Journal of Communications}\
  }\textbf {\bibinfo {volume} {5}},\ \bibinfo {pages} {1375} (\bibinfo {year}
  {2011})}\BibitemShut {NoStop}%
\bibitem [{\citenamefont {Farrell}(2012)}]{Farrell2012}%
  \BibitemOpen
  \bibfield  {author} {\bibinfo {author} {\bibfnamefont {H.}~\bibnamefont
  {Farrell}},\ }\href {\doibase 10.1146/annurev-polisci-030810-110815}
  {\bibfield  {journal} {\bibinfo  {journal} {Annual Review of Political
  Science}\ }\textbf {\bibinfo {volume} {15}},\ \bibinfo {pages} {35} (\bibinfo
  {year} {2012})}\BibitemShut {NoStop}%
\bibitem [{\citenamefont {Barnsby}(2012)}]{Barnsby2012}%
  \BibitemOpen
  \bibfield  {author} {\bibinfo {author} {\bibfnamefont {R.~E.}\ \bibnamefont
  {Barnsby}},\ }\emph {\bibinfo {title} {Social Media and the Arab Spring: How
  Facebook, Twitter, and Camera Phones Changed the Egyptian Army's Response to
  Revolution}},\ \href@noop {} {Master's thesis} (\bibinfo {year}
  {2012})\BibitemShut {NoStop}%
\bibitem [{\citenamefont {Zhao}\ \emph {et~al.}(2010)\citenamefont {Zhao},
  \citenamefont {Wu},\ and\ \citenamefont {Xu}}]{WeakTies-PRE-2010}%
  \BibitemOpen
  \bibfield  {author} {\bibinfo {author} {\bibfnamefont {J.}~\bibnamefont
  {Zhao}}, \bibinfo {author} {\bibfnamefont {J.}~\bibnamefont {Wu}}, \ and\
  \bibinfo {author} {\bibfnamefont {K.}~\bibnamefont {Xu}},\ }\href {\doibase
  10.1103/PhysRevE.82.016105} {\bibfield  {journal} {\bibinfo  {journal} {Phys.
  Rev. E}\ }\textbf {\bibinfo {volume} {82}},\ \bibinfo {pages} {016105}
  (\bibinfo {year} {2010})}\BibitemShut {NoStop}%
\bibitem [{\citenamefont {G\'omez}\ \emph {et~al.}(2013)\citenamefont
  {G\'omez}, \citenamefont {D\'iaz-Guilera}, \citenamefont
  {G\'omez-Garde\~nes}, \citenamefont {P\'erez-Vicente}, \citenamefont
  {Moreno},\ and\ \citenamefont {Arenas}}]{GomezEtAl-PRL-2013}%
  \BibitemOpen
  \bibfield  {author} {\bibinfo {author} {\bibfnamefont {S.}~\bibnamefont
  {G\'omez}}, \bibinfo {author} {\bibfnamefont {A.}~\bibnamefont
  {D\'iaz-Guilera}}, \bibinfo {author} {\bibfnamefont {J.}~\bibnamefont
  {G\'omez-Garde\~nes}}, \bibinfo {author} {\bibfnamefont {C.~J.}\ \bibnamefont
  {P\'erez-Vicente}}, \bibinfo {author} {\bibfnamefont {Y.}~\bibnamefont
  {Moreno}}, \ and\ \bibinfo {author} {\bibfnamefont {A.}~\bibnamefont
  {Arenas}},\ }\href {\doibase 10.1103/PhysRevLett.110.028701} {\bibfield
  {journal} {\bibinfo  {journal} {Phys. Rev. Lett.}\ }\textbf {\bibinfo
  {volume} {110}},\ \bibinfo {pages} {028701} (\bibinfo {year}
  {2013})}\BibitemShut {NoStop}%
\bibitem [{\citenamefont {Myers}\ \emph {et~al.}(2012)\citenamefont {Myers},
  \citenamefont {Zhu},\ and\ \citenamefont {Leskovec}}]{MyersEtAl-arxiv-2012}%
  \BibitemOpen
  \bibfield  {author} {\bibinfo {author} {\bibfnamefont {S.~A.}\ \bibnamefont
  {Myers}}, \bibinfo {author} {\bibfnamefont {C.}~\bibnamefont {Zhu}}, \ and\
  \bibinfo {author} {\bibfnamefont {J.}~\bibnamefont {Leskovec}},\ }\href@noop
  {} {\bibfield  {journal} {\bibinfo  {journal} {arXiv}\ }\textbf {\bibinfo
  {volume} {http://arxiv.org/abs/1206.1331}} (\bibinfo {year}
  {2012})}\BibitemShut {NoStop}%
\bibitem [{\citenamefont {Guille}\ \emph {et~al.}(2013)\citenamefont {Guille},
  \citenamefont {Hacid},\ and\ \citenamefont {Favre}}]{GuilleEtAl-arxiv-2013}%
  \BibitemOpen
  \bibfield  {author} {\bibinfo {author} {\bibfnamefont {A.}~\bibnamefont
  {Guille}}, \bibinfo {author} {\bibfnamefont {H.}~\bibnamefont {Hacid}}, \
  and\ \bibinfo {author} {\bibfnamefont {C.}~\bibnamefont {Favre}},\
  }\href@noop {} {\bibfield  {journal} {\bibinfo  {journal} {arXiv}\ }\textbf
  {\bibinfo {volume} {http://arxiv.org/abs/1302.5235v2}} (\bibinfo {year}
  {2013})}\BibitemShut {NoStop}%
\bibitem [{\citenamefont {Shafiq}\ and\ \citenamefont
  {Liu}(2013)}]{social-cascades-arxiv-2013}%
  \BibitemOpen
  \bibfield  {author} {\bibinfo {author} {\bibfnamefont {M.~Z.}\ \bibnamefont
  {Shafiq}}\ and\ \bibinfo {author} {\bibfnamefont {A.~X.}\ \bibnamefont
  {Liu}},\ }\href@noop {} {\bibfield  {journal} {\bibinfo  {journal} {arXiv}\
  }\textbf {\bibinfo {volume} {http://arxiv.org/abs/1302.2376}} (\bibinfo
  {year} {2013})}\BibitemShut {NoStop}%
\bibitem [{\citenamefont {Ba\~{n}os}\ \emph {et~al.}(2013)\citenamefont
  {Ba\~{n}os}, \citenamefont {Borge-Holthoefer},\ and\ \citenamefont
  {Moreno}}]{info-cascades-arxiv-2013}%
  \BibitemOpen
  \bibfield  {author} {\bibinfo {author} {\bibfnamefont {R.~A.}\ \bibnamefont
  {Ba\~{n}os}}, \bibinfo {author} {\bibfnamefont {J.}~\bibnamefont
  {Borge-Holthoefer}}, \ and\ \bibinfo {author} {\bibfnamefont
  {Y.}~\bibnamefont {Moreno}},\ }\href@noop {} {\bibfield  {journal} {\bibinfo
  {journal} {arXiv}\ }\textbf {\bibinfo {volume}
  {http://arxiv.org/abs/1303.4629}} (\bibinfo {year} {2013})}\BibitemShut
  {NoStop}%
\bibitem [{\citenamefont {Leskovec}\ \emph {et~al.}(2009)\citenamefont
  {Leskovec}, \citenamefont {Backstrom},\ and\ \citenamefont
  {Kleinberg}}]{Leskovec2009}%
  \BibitemOpen
  \bibfield  {author} {\bibinfo {author} {\bibfnamefont {J.}~\bibnamefont
  {Leskovec}}, \bibinfo {author} {\bibfnamefont {L.}~\bibnamefont {Backstrom}},
  \ and\ \bibinfo {author} {\bibfnamefont {J.}~\bibnamefont {Kleinberg}},\
  }\href@noop {} {\bibfield  {journal} {\bibinfo  {journal} {Proc 15th ACM
  SIGKDD Int Conf on Knowledge Discovery and Data Mining}\ ,\ \bibinfo {pages}
  {497–505}} (\bibinfo {year} {2009})}\BibitemShut {NoStop}%
\bibitem [{\citenamefont {Leskovec}\ \emph {et~al.}(2007)\citenamefont
  {Leskovec}, \citenamefont {McGlohon}, \citenamefont {Faloutsos},
  \citenamefont {Glance},\ and\ \citenamefont {Hurst}}]{Leskovec2007}%
  \BibitemOpen
  \bibfield  {author} {\bibinfo {author} {\bibfnamefont {J.}~\bibnamefont
  {Leskovec}}, \bibinfo {author} {\bibfnamefont {M.}~\bibnamefont {McGlohon}},
  \bibinfo {author} {\bibfnamefont {C.}~\bibnamefont {Faloutsos}}, \bibinfo
  {author} {\bibfnamefont {N.}~\bibnamefont {Glance}}, \ and\ \bibinfo {author}
  {\bibfnamefont {M.}~\bibnamefont {Hurst}},\ }\href@noop {} {\bibfield
  {journal} {\bibinfo  {journal} {Proc. SIAM Int. Conf. on Data Mining}\ ,\
  \bibinfo {pages} {551–556}} (\bibinfo {year} {2007})}\BibitemShut {NoStop}%
\bibitem [{\citenamefont {Onnela}\ and\ \citenamefont
  {Reed-Tsochas}(2010)}]{Onnela2010}%
  \BibitemOpen
  \bibfield  {author} {\bibinfo {author} {\bibfnamefont {J.-P.}\ \bibnamefont
  {Onnela}}\ and\ \bibinfo {author} {\bibfnamefont {F.}~\bibnamefont
  {Reed-Tsochas}},\ }\href@noop {} {\bibfield  {journal} {\bibinfo  {journal}
  {PNAS}\ }\textbf {\bibinfo {volume} {107}} (\bibinfo {year}
  {2010})}\BibitemShut {NoStop}%
\bibitem [{\citenamefont {Bakshy}\ \emph {et~al.}(2011)\citenamefont {Bakshy},
  \citenamefont {Hofman}, \citenamefont {Mason},\ and\ \citenamefont
  {Watts}}]{Bakshy2011}%
  \BibitemOpen
  \bibfield  {author} {\bibinfo {author} {\bibfnamefont {E.}~\bibnamefont
  {Bakshy}}, \bibinfo {author} {\bibfnamefont {J.}~\bibnamefont {Hofman}},
  \bibinfo {author} {\bibfnamefont {W.}~\bibnamefont {Mason}}, \ and\ \bibinfo
  {author} {\bibfnamefont {D.}~\bibnamefont {Watts}},\ }\href@noop {}
  {\bibfield  {journal} {\bibinfo  {journal} {Proc. 4th Int. Conf. on Web
  Search and Data Mining}\ ,\ \bibinfo {pages} {65–74}} (\bibinfo {year}
  {2011})}\BibitemShut {NoStop}%
\bibitem [{\citenamefont {Lerman}\ and\ \citenamefont
  {Ghosh}(2010)}]{LermanGhosh2010}%
  \BibitemOpen
  \bibfield  {author} {\bibinfo {author} {\bibfnamefont {K.}~\bibnamefont
  {Lerman}}\ and\ \bibinfo {author} {\bibfnamefont {R.}~\bibnamefont {Ghosh}},\
  }in\ \href@noop {} {\emph {\bibinfo {booktitle} {Proceedings of the Fourth
  International AAAI Conference on Weblogs and Social Media}}}\ (\bibinfo
  {organization} {The Association for the Advancement of Artificial
  Intelligence},\ \bibinfo {year} {2010})\ pp.\ \bibinfo {pages}
  {90--97}\BibitemShut {NoStop}%
\bibitem [{\citenamefont {Simmons}\ \emph {et~al.}(2011)\citenamefont
  {Simmons}, \citenamefont {Adamic},\ and\ \citenamefont {Adar}}]{Simmons2011}%
  \BibitemOpen
  \bibfield  {author} {\bibinfo {author} {\bibfnamefont {M.~P.}\ \bibnamefont
  {Simmons}}, \bibinfo {author} {\bibfnamefont {L.~A.}\ \bibnamefont {Adamic}},
  \ and\ \bibinfo {author} {\bibfnamefont {E.}~\bibnamefont {Adar}},\
  }\href@noop {} {\bibfield  {journal} {\bibinfo  {journal} {Proc. 5th Int.
  AAAI Conf. on Weblogs and Social Media}\ ,\ \bibinfo {pages} {353–360}}
  (\bibinfo {year} {2011})}\BibitemShut {NoStop}%
\bibitem [{\citenamefont {Watts}(2002)}]{Watts2002}%
  \BibitemOpen
  \bibfield  {author} {\bibinfo {author} {\bibfnamefont {D.~J.}\ \bibnamefont
  {Watts}},\ }\href@noop {} {\bibfield  {journal} {\bibinfo  {journal} {Proc
  Natl Acad Sci U S A}\ }\textbf {\bibinfo {volume} {99}},\ \bibinfo {pages}
  {5766} (\bibinfo {year} {2002})}\BibitemShut {NoStop}%
\bibitem [{\citenamefont {Doerr}\ \emph {et~al.}(2012)\citenamefont {Doerr},
  \citenamefont {Fouz},\ and\ \citenamefont {Friedrich}}]{Doerr2012}%
  \BibitemOpen
  \bibfield  {author} {\bibinfo {author} {\bibfnamefont {B.}~\bibnamefont
  {Doerr}}, \bibinfo {author} {\bibfnamefont {M.}~\bibnamefont {Fouz}}, \ and\
  \bibinfo {author} {\bibfnamefont {T.}~\bibnamefont {Friedrich}},\ }\href
  {\doibase 10.1145/2184319.2184338} {\bibfield  {journal} {\bibinfo  {journal}
  {Commun. ACM}\ }\textbf {\bibinfo {volume} {55}},\ \bibinfo {pages} {70}
  (\bibinfo {year} {2012})}\BibitemShut {NoStop}%
\bibitem [{\citenamefont {Zhang}\ \emph {et~al.}(2013)\citenamefont {Zhang},
  \citenamefont {Zhou}, \citenamefont {Zhang}, \citenamefont {Guan},\ and\
  \citenamefont {Zhou}}]{ZhangEtAl-PRE-2013}%
  \BibitemOpen
  \bibfield  {author} {\bibinfo {author} {\bibfnamefont {Y.}~\bibnamefont
  {Zhang}}, \bibinfo {author} {\bibfnamefont {S.}~\bibnamefont {Zhou}},
  \bibinfo {author} {\bibfnamefont {Z.}~\bibnamefont {Zhang}}, \bibinfo
  {author} {\bibfnamefont {J.}~\bibnamefont {Guan}}, \ and\ \bibinfo {author}
  {\bibfnamefont {S.}~\bibnamefont {Zhou}},\ }\href {\doibase
  10.1103/PhysRevE.87.032133} {\bibfield  {journal} {\bibinfo  {journal} {Phys.
  Rev. E}\ }\textbf {\bibinfo {volume} {87}},\ \bibinfo {pages} {032133}
  (\bibinfo {year} {2013})}\BibitemShut {NoStop}%
\bibitem [{\citenamefont {Kitsak}\ \emph {et~al.}(2010)\citenamefont {Kitsak},
  \citenamefont {Gallos}, \citenamefont {Havlin}, \citenamefont {Liljeros},
  \citenamefont {Muchnik}, \citenamefont {Stanley},\ and\ \citenamefont
  {Makse}}]{SpreadersNature2010}%
  \BibitemOpen
  \bibfield  {author} {\bibinfo {author} {\bibfnamefont {M.}~\bibnamefont
  {Kitsak}}, \bibinfo {author} {\bibfnamefont {L.~K.}\ \bibnamefont {Gallos}},
  \bibinfo {author} {\bibfnamefont {S.}~\bibnamefont {Havlin}}, \bibinfo
  {author} {\bibfnamefont {F.}~\bibnamefont {Liljeros}}, \bibinfo {author}
  {\bibfnamefont {L.}~\bibnamefont {Muchnik}}, \bibinfo {author} {\bibfnamefont
  {H.~E.}\ \bibnamefont {Stanley}}, \ and\ \bibinfo {author} {\bibfnamefont
  {H.~A.}\ \bibnamefont {Makse}},\ }\href@noop {} {\bibfield  {journal}
  {\bibinfo  {journal} {Nature Physics}\ }\textbf {\bibinfo {volume} {6}},\
  \bibinfo {pages} {888} (\bibinfo {year} {2010})}\BibitemShut {NoStop}%
\bibitem [{\citenamefont {Borge-Holthoefer}\ and\ \citenamefont
  {Moreno}(2012)}]{SpreaderAbsence-PRE-2012}%
  \BibitemOpen
  \bibfield  {author} {\bibinfo {author} {\bibfnamefont {J.}~\bibnamefont
  {Borge-Holthoefer}}\ and\ \bibinfo {author} {\bibfnamefont {Y.}~\bibnamefont
  {Moreno}},\ }\href {\doibase 10.1103/PhysRevE.85.026116} {\bibfield
  {journal} {\bibinfo  {journal} {Phys. Rev. E}\ }\textbf {\bibinfo {volume}
  {85}},\ \bibinfo {pages} {026116} (\bibinfo {year} {2012})}\BibitemShut
  {NoStop}%
\bibitem [{\citenamefont {Borge-Holthoefer}\ \emph {et~al.}(2012)\citenamefont
  {Borge-Holthoefer}, \citenamefont {Rivero},\ and\ \citenamefont
  {Moreno}}]{FindingSpreaders-PRE-2012}%
  \BibitemOpen
  \bibfield  {author} {\bibinfo {author} {\bibfnamefont {J.}~\bibnamefont
  {Borge-Holthoefer}}, \bibinfo {author} {\bibfnamefont {A.}~\bibnamefont
  {Rivero}}, \ and\ \bibinfo {author} {\bibfnamefont {Y.}~\bibnamefont
  {Moreno}},\ }\href {\doibase 10.1103/PhysRevE.85.066123} {\bibfield
  {journal} {\bibinfo  {journal} {Phys. Rev. E}\ }\textbf {\bibinfo {volume}
  {85}},\ \bibinfo {pages} {066123} (\bibinfo {year} {2012})}\BibitemShut
  {NoStop}%
\bibitem [{\citenamefont {Centola}(2011)}]{Centola2011}%
  \BibitemOpen
  \bibfield  {author} {\bibinfo {author} {\bibfnamefont {D.}~\bibnamefont
  {Centola}},\ }\href@noop {} {\bibfield  {journal} {\bibinfo  {journal}
  {Science}\ }\textbf {\bibinfo {volume} {334}},\ \bibinfo {pages} {1269}
  (\bibinfo {year} {2011})}\BibitemShut {NoStop}%
\bibitem [{\citenamefont {Dodds}\ and\ \citenamefont
  {Watts}(2005)}]{Dodds2005}%
  \BibitemOpen
  \bibfield  {author} {\bibinfo {author} {\bibfnamefont {P.~S.}\ \bibnamefont
  {Dodds}}\ and\ \bibinfo {author} {\bibfnamefont {D.~J.}\ \bibnamefont
  {Watts}},\ }\href@noop {} {\bibfield  {journal} {\bibinfo  {journal} {J Theor
  Biol}\ }\textbf {\bibinfo {volume} {232}},\ \bibinfo {pages} {587–604}
  (\bibinfo {year} {2005})}\BibitemShut {NoStop}%
\bibitem [{\citenamefont {Bettencourt}\ \emph {et~al.}(2006)\citenamefont
  {Bettencourt}, \citenamefont {Cintron-Arias}, \citenamefont {Kaiser},\ and\
  \citenamefont {Castillo-Chavez}}]{Bettencourt2006}%
  \BibitemOpen
  \bibfield  {author} {\bibinfo {author} {\bibfnamefont {L.~M.~A.}\
  \bibnamefont {Bettencourt}}, \bibinfo {author} {\bibfnamefont
  {A.}~\bibnamefont {Cintron-Arias}}, \bibinfo {author} {\bibfnamefont {D.~I.}\
  \bibnamefont {Kaiser}}, \ and\ \bibinfo {author} {\bibfnamefont
  {C.}~\bibnamefont {Castillo-Chavez}},\ }\href@noop {} {\bibfield  {journal}
  {\bibinfo  {journal} {Physica A}\ }\textbf {\bibinfo {volume} {364}},\
  \bibinfo {pages} {513} (\bibinfo {year} {2006})}\BibitemShut {NoStop}%
\bibitem [{\citenamefont {Diaz-Aviles}\ \emph {et~al.}(2012)\citenamefont
  {Diaz-Aviles}, \citenamefont {Stewart}, \citenamefont {Velasco},
  \citenamefont {Denecke},\ and\ \citenamefont
  {Nejdl}}]{epidemic-intelligence-arxiv-2012}%
  \BibitemOpen
  \bibfield  {author} {\bibinfo {author} {\bibfnamefont {E.}~\bibnamefont
  {Diaz-Aviles}}, \bibinfo {author} {\bibfnamefont {A.}~\bibnamefont
  {Stewart}}, \bibinfo {author} {\bibfnamefont {E.}~\bibnamefont {Velasco}},
  \bibinfo {author} {\bibfnamefont {K.}~\bibnamefont {Denecke}}, \ and\
  \bibinfo {author} {\bibfnamefont {W.}~\bibnamefont {Nejdl}},\ }\href@noop {}
  {\bibfield  {journal} {\bibinfo  {journal} {arXiv}\ }\textbf {\bibinfo
  {volume} {http://arxiv.org/abs/1203.1378}} (\bibinfo {year}
  {2012})}\BibitemShut {NoStop}%
\bibitem [{\citenamefont {Li}\ \emph {et~al.}(2013)\citenamefont {Li},
  \citenamefont {Wang},\ and\ \citenamefont
  {Van~Mieghem}}]{epidemic-threshold-arxiv-2013}%
  \BibitemOpen
  \bibfield  {author} {\bibinfo {author} {\bibfnamefont {C.}~\bibnamefont
  {Li}}, \bibinfo {author} {\bibfnamefont {H.}~\bibnamefont {Wang}}, \ and\
  \bibinfo {author} {\bibfnamefont {P.}~\bibnamefont {Van~Mieghem}},\
  }\href@noop {} {\bibfield  {journal} {\bibinfo  {journal} {arXiv}\ }\textbf
  {\bibinfo {volume} {http://arxiv.org/abs/1303.0783}} (\bibinfo {year}
  {2013})}\BibitemShut {NoStop}%
\bibitem [{\citenamefont {Burt}(1987)}]{Burt1987}%
  \BibitemOpen
  \bibfield  {author} {\bibinfo {author} {\bibfnamefont {R.~S.}\ \bibnamefont
  {Burt}},\ }\href@noop {} {\bibfield  {journal} {\bibinfo  {journal} {American
  Journal of Sociology}\ }\textbf {\bibinfo {volume} {92}},\ \bibinfo {pages}
  {1287} (\bibinfo {year} {1987})}\BibitemShut {NoStop}%
\bibitem [{\citenamefont {Melnik}\ \emph {et~al.}(2013)\citenamefont {Melnik},
  \citenamefont {Ward}, \citenamefont {Gleeson},\ and\ \citenamefont
  {Porter}}]{Melnik2011}%
  \BibitemOpen
  \bibfield  {author} {\bibinfo {author} {\bibfnamefont {S.}~\bibnamefont
  {Melnik}}, \bibinfo {author} {\bibfnamefont {J.~A.}\ \bibnamefont {Ward}},
  \bibinfo {author} {\bibfnamefont {J.~P.}\ \bibnamefont {Gleeson}}, \ and\
  \bibinfo {author} {\bibfnamefont {M.~A.}\ \bibnamefont {Porter}},\
  }\href@noop {} {\bibfield  {journal} {\bibinfo  {journal} {Chaos}\ }\textbf
  {\bibinfo {volume} {23}},\ \bibinfo {pages} {013124} (\bibinfo {year}
  {2013})}\BibitemShut {NoStop}%
\bibitem [{\citenamefont {Centola}\ \emph {et~al.}(2007)\citenamefont
  {Centola}, \citenamefont {Egu\'{\i}luz},\ and\ \citenamefont
  {Macy}}]{Centola2007}%
  \BibitemOpen
  \bibfield  {author} {\bibinfo {author} {\bibfnamefont {D.}~\bibnamefont
  {Centola}}, \bibinfo {author} {\bibfnamefont {V.~M.}\ \bibnamefont
  {Egu\'{\i}luz}}, \ and\ \bibinfo {author} {\bibfnamefont {M.~W.}\
  \bibnamefont {Macy}},\ }\href@noop {} {\bibfield  {journal} {\bibinfo
  {journal} {Physica A}\ }\textbf {\bibinfo {volume} {374}},\ \bibinfo {pages}
  {449} (\bibinfo {year} {2007})}\BibitemShut {NoStop}%
\bibitem [{\citenamefont {Centola}\ and\ \citenamefont
  {Macy}(2007)}]{Centola2007b}%
  \BibitemOpen
  \bibfield  {author} {\bibinfo {author} {\bibfnamefont {D.}~\bibnamefont
  {Centola}}\ and\ \bibinfo {author} {\bibfnamefont {M.}~\bibnamefont {Macy}},\
  }\href@noop {} {\bibfield  {journal} {\bibinfo  {journal} {American Journal
  of Sociology}\ }\textbf {\bibinfo {volume} {113}},\ \bibinfo {pages} {702}
  (\bibinfo {year} {2007})}\BibitemShut {NoStop}%
\bibitem [{\citenamefont {Centola}(2010)}]{Centola2010}%
  \BibitemOpen
  \bibfield  {author} {\bibinfo {author} {\bibfnamefont {D.}~\bibnamefont
  {Centola}},\ }\href@noop {} {\bibfield  {journal} {\bibinfo  {journal}
  {Science}\ }\textbf {\bibinfo {volume} {329}},\ \bibinfo {pages} {1194}
  (\bibinfo {year} {2010})}\BibitemShut {NoStop}%
\bibitem [{\citenamefont {Bassett}\ \emph {et~al.}(2012)\citenamefont
  {Bassett}, \citenamefont {Alderson},\ and\ \citenamefont
  {Carlson}}]{BassettEtAl2012-PRE}%
  \BibitemOpen
  \bibfield  {author} {\bibinfo {author} {\bibfnamefont {D.}~\bibnamefont
  {Bassett}}, \bibinfo {author} {\bibfnamefont {D.}~\bibnamefont {Alderson}}, \
  and\ \bibinfo {author} {\bibfnamefont {J.}~\bibnamefont {Carlson}},\
  }\href@noop {} {\bibfield  {journal} {\bibinfo  {journal} {Physical Review
  E}\ }\textbf {\bibinfo {volume} {86}} (\bibinfo {year} {2012})}\BibitemShut
  {NoStop}%
\bibitem [{\citenamefont {Barahona}\ \emph {et~al.}(2012)\citenamefont
  {Barahona}, \citenamefont {Garc'a}, \citenamefont {Gloor},\ and\
  \citenamefont {Parraguez}}]{BarahonaEtAl2012}%
  \BibitemOpen
  \bibfield  {author} {\bibinfo {author} {\bibfnamefont {M.}~\bibnamefont
  {Barahona}}, \bibinfo {author} {\bibfnamefont {C.}~\bibnamefont {Garc'a}},
  \bibinfo {author} {\bibfnamefont {P.}~\bibnamefont {Gloor}}, \ and\ \bibinfo
  {author} {\bibfnamefont {P.}~\bibnamefont {Parraguez}},\ }in\ \href@noop {}
  {\emph {\bibinfo {booktitle} {Proceedings of the 2012 Conference on
  Collective Intelligence}}}\ (\bibinfo {year} {2012})\BibitemShut {NoStop}%
\bibitem [{\citenamefont {Sano}\ \emph {et~al.}(2013)\citenamefont {Sano},
  \citenamefont {Yamada}, \citenamefont {Watanabe}, \citenamefont {Takayasu},\
  and\ \citenamefont {Takayasu}}]{SanoEtAl-PRE2013}%
  \BibitemOpen
  \bibfield  {author} {\bibinfo {author} {\bibfnamefont {Y.}~\bibnamefont
  {Sano}}, \bibinfo {author} {\bibfnamefont {K.}~\bibnamefont {Yamada}},
  \bibinfo {author} {\bibfnamefont {H.}~\bibnamefont {Watanabe}}, \bibinfo
  {author} {\bibfnamefont {H.}~\bibnamefont {Takayasu}}, \ and\ \bibinfo
  {author} {\bibfnamefont {M.}~\bibnamefont {Takayasu}},\ }\href {\doibase
  10.1103/PhysRevE.87.012805} {\bibfield  {journal} {\bibinfo  {journal} {Phys.
  Rev. E}\ }\textbf {\bibinfo {volume} {87}},\ \bibinfo {pages} {012805}
  (\bibinfo {year} {2013})}\BibitemShut {NoStop}%
\bibitem [{\citenamefont {Kearns}\ \emph {et~al.}(2006)\citenamefont {Kearns},
  \citenamefont {Suri},\ and\ \citenamefont {Montfort}}]{Kearns2006}%
  \BibitemOpen
  \bibfield  {author} {\bibinfo {author} {\bibfnamefont {M.}~\bibnamefont
  {Kearns}}, \bibinfo {author} {\bibfnamefont {S.}~\bibnamefont {Suri}}, \ and\
  \bibinfo {author} {\bibfnamefont {N.}~\bibnamefont {Montfort}},\ }\href@noop
  {} {\bibfield  {journal} {\bibinfo  {journal} {Science}\ }\textbf {\bibinfo
  {volume} {313}},\ \bibinfo {pages} {824} (\bibinfo {year}
  {2006})}\BibitemShut {NoStop}%
\bibitem [{\citenamefont {Kearns}\ \emph {et~al.}(2009)\citenamefont {Kearns},
  \citenamefont {Judd}, \citenamefont {Tan},\ and\ \citenamefont
  {Wortman}}]{Kearns2009}%
  \BibitemOpen
  \bibfield  {author} {\bibinfo {author} {\bibfnamefont {M.}~\bibnamefont
  {Kearns}}, \bibinfo {author} {\bibfnamefont {S.}~\bibnamefont {Judd}},
  \bibinfo {author} {\bibfnamefont {J.}~\bibnamefont {Tan}}, \ and\ \bibinfo
  {author} {\bibfnamefont {J.}~\bibnamefont {Wortman}},\ }\href@noop {}
  {\bibfield  {journal} {\bibinfo  {journal} {Proc Natl Acad Sci U S A}\
  }\textbf {\bibinfo {volume} {106}},\ \bibinfo {pages} {1347} (\bibinfo {year}
  {2009})}\BibitemShut {NoStop}%
\bibitem [{\citenamefont {Judd}\ \emph {et~al.}(2010)\citenamefont {Judd},
  \citenamefont {Kearns},\ and\ \citenamefont {Vorobeychik}}]{Kearns2010}%
  \BibitemOpen
  \bibfield  {author} {\bibinfo {author} {\bibfnamefont {S.}~\bibnamefont
  {Judd}}, \bibinfo {author} {\bibfnamefont {M.}~\bibnamefont {Kearns}}, \ and\
  \bibinfo {author} {\bibfnamefont {Y.}~\bibnamefont {Vorobeychik}},\
  }\href@noop {} {\bibfield  {journal} {\bibinfo  {journal} {Proc Natl Acad Sci
  U S A}\ }\textbf {\bibinfo {volume} {108}},\ \bibinfo {pages} {6685}
  (\bibinfo {year} {2010})}\BibitemShut {NoStop}%
\bibitem [{\citenamefont {Kearns}(2012)}]{Kearns2012}%
  \BibitemOpen
  \bibfield  {author} {\bibinfo {author} {\bibfnamefont {M.}~\bibnamefont
  {Kearns}},\ }\href@noop {} {\bibfield  {journal} {\bibinfo  {journal}
  {Communications of the ACM}\ }\textbf {\bibinfo {volume} {55}},\ \bibinfo
  {pages} {56} (\bibinfo {year} {2012})}\BibitemShut {NoStop}%
\bibitem [{\citenamefont {Mason}\ and\ \citenamefont
  {Watts}(2012)}]{mason_collaborative_2012}%
  \BibitemOpen
  \bibfield  {author} {\bibinfo {author} {\bibfnamefont {W.}~\bibnamefont
  {Mason}}\ and\ \bibinfo {author} {\bibfnamefont {D.~J.}\ \bibnamefont
  {Watts}},\ }\href {\doibase 10.1073/pnas.1110069108} {\bibfield  {journal}
  {\bibinfo  {journal} {Proceedings of the National Academy of Sciences of the
  United States of America}\ }\textbf {\bibinfo {volume} {109}},\ \bibinfo
  {pages} {764} (\bibinfo {year} {2012})}\BibitemShut {NoStop}%
\bibitem [{\citenamefont {Nedic}\ \emph {et~al.}(2012)\citenamefont {Nedic},
  \citenamefont {Tomlin}, \citenamefont {Holmes}, \citenamefont {Prentice},\
  and\ \citenamefont {Cohen}}]{nedic_decision_2012}%
  \BibitemOpen
  \bibfield  {author} {\bibinfo {author} {\bibfnamefont {A.}~\bibnamefont
  {Nedic}}, \bibinfo {author} {\bibfnamefont {D.}~\bibnamefont {Tomlin}},
  \bibinfo {author} {\bibfnamefont {P.}~\bibnamefont {Holmes}}, \bibinfo
  {author} {\bibfnamefont {D.}~\bibnamefont {Prentice}}, \ and\ \bibinfo
  {author} {\bibfnamefont {J.~D.}\ \bibnamefont {Cohen}},\ }\href {\doibase
  10.1109/JPROC.2011.2166437} {\bibfield  {journal} {\bibinfo  {journal}
  {Proceedings of the {IEEE}}\ }\textbf {\bibinfo {volume} {100}},\ \bibinfo
  {pages} {713} (\bibinfo {year} {2012})}\BibitemShut {NoStop}%
\bibitem [{\citenamefont {Drabek}(1986)}]{Drabek1986}%
  \BibitemOpen
  \bibfield  {author} {\bibinfo {author} {\bibfnamefont {T.~E.}\ \bibnamefont
  {Drabek}},\ }\href@noop {} {\emph {\bibinfo {title} {Human systems responses
  to disaster: An inventory of sociological findings}}}\ (\bibinfo  {publisher}
  {Springer},\ \bibinfo {year} {1986})\ p.~\bibinfo {pages} {74}\BibitemShut
  {NoStop}%
\bibitem [{\citenamefont {Lindell}\ \emph {et~al.}(2006)\citenamefont
  {Lindell}, \citenamefont {Prater},\ and\ \citenamefont {Perry}}]{LPP06}%
  \BibitemOpen
  \bibfield  {author} {\bibinfo {author} {\bibfnamefont {M.~K.}\ \bibnamefont
  {Lindell}}, \bibinfo {author} {\bibfnamefont {C.}~\bibnamefont {Prater}}, \
  and\ \bibinfo {author} {\bibfnamefont {R.~W.}\ \bibnamefont {Perry}},\
  }\href@noop {} {\emph {\bibinfo {title} {Emergency Management}}}\ (\bibinfo
  {publisher} {Wiley},\ \bibinfo {year} {2006})\BibitemShut {NoStop}%
\bibitem [{\citenamefont {Dash}\ and\ \citenamefont {Gladwin}(2007)}]{DH07}%
  \BibitemOpen
  \bibfield  {author} {\bibinfo {author} {\bibfnamefont {N.}~\bibnamefont
  {Dash}}\ and\ \bibinfo {author} {\bibfnamefont {H.}~\bibnamefont {Gladwin}},\
  }\href@noop {} {\bibfield  {journal} {\bibinfo  {journal} {Natural Hazards
  Review}\ }\textbf {\bibinfo {volume} {8}},\ \bibinfo {pages} {69} (\bibinfo
  {year} {2007})}\BibitemShut {NoStop}%
\bibitem [{\citenamefont {Sweeney}(2008)}]{Sweeney2008}%
  \BibitemOpen
  \bibfield  {author} {\bibinfo {author} {\bibfnamefont {K.}~\bibnamefont
  {Sweeney}},\ }\href@noop {} {\bibfield  {journal} {\bibinfo  {journal}
  {Psychological Bulletin}\ }\textbf {\bibinfo {volume} {134}},\ \bibinfo
  {pages} {61} (\bibinfo {year} {2008})}\BibitemShut {NoStop}%
\bibitem [{\citenamefont {Conneally}()}]{Conneally2011}%
  \BibitemOpen
  \bibfield  {author} {\bibinfo {author} {\bibfnamefont {T.}~\bibnamefont
  {Conneally}},\ }\href@noop {} {\enquote {\bibinfo {title} {Virginia
  earthquake overloads cell networks from {N}orth {C}arolina to {N}ew {Y}ork,
  {T}witter takes over},}\ }\bibinfo {howpublished}
  {{http://betanews.com/2011/08/23/virginia-earthquake-knocks-out-regions-cell-networks/}},\
  \bibinfo {note} {published: 23 August 2011. Last accessed: 2 January
  2012.}\BibitemShut {Stop}%
\bibitem [{\citenamefont {Helbing}\ \emph {et~al.}(2000)\citenamefont
  {Helbing}, \citenamefont {Farkas},\ and\ \citenamefont
  {Vicsek}}]{helbing_simulating_2000}%
  \BibitemOpen
  \bibfield  {author} {\bibinfo {author} {\bibfnamefont {D.}~\bibnamefont
  {Helbing}}, \bibinfo {author} {\bibfnamefont {I.}~\bibnamefont {Farkas}}, \
  and\ \bibinfo {author} {\bibfnamefont {T.}~\bibnamefont {Vicsek}},\ }\href
  {\doibase 10.1038/35035023} {\bibfield  {journal} {\bibinfo  {journal}
  {Nature}\ }\textbf {\bibinfo {volume} {407}},\ \bibinfo {pages} {487}
  (\bibinfo {year} {2000})}\BibitemShut {NoStop}%
\bibitem [{\citenamefont {Helbing}(2001)}]{helbing_traffic_2001}%
  \BibitemOpen
  \bibfield  {author} {\bibinfo {author} {\bibfnamefont {D.}~\bibnamefont
  {Helbing}},\ }\href {\doibase 10.1103/RevModPhys.73.1067} {\bibfield
  {journal} {\bibinfo  {journal} {Reviews of Modern Physics}\ }\textbf
  {\bibinfo {volume} {73}},\ \bibinfo {pages} {1067} (\bibinfo {year}
  {2001})}\BibitemShut {NoStop}%
\bibitem [{\citenamefont {Banerjee}(1992)}]{Banerjee1992}%
  \BibitemOpen
  \bibfield  {author} {\bibinfo {author} {\bibfnamefont {A.}~\bibnamefont
  {Banerjee}},\ }\href@noop {} {\bibfield  {journal} {\bibinfo  {journal}
  {Quarterly Journal of Economics}\ }\textbf {\bibinfo {volume} {107}},\
  \bibinfo {pages} {797} (\bibinfo {year} {1992})}\BibitemShut {NoStop}%
\bibitem [{\citenamefont {Bosse}\ \emph {et~al.}(2012)\citenamefont {Bosse},
  \citenamefont {Hoogendoorn}, \citenamefont {Klein}, \citenamefont {Treur},
  \citenamefont {van derWal},\ and\ \citenamefont {vanWissen}}]{Bosse2012}%
  \BibitemOpen
  \bibfield  {author} {\bibinfo {author} {\bibfnamefont {T.}~\bibnamefont
  {Bosse}}, \bibinfo {author} {\bibfnamefont {M.}~\bibnamefont {Hoogendoorn}},
  \bibinfo {author} {\bibfnamefont {M.~C.~A.}\ \bibnamefont {Klein}}, \bibinfo
  {author} {\bibfnamefont {J.}~\bibnamefont {Treur}}, \bibinfo {author}
  {\bibfnamefont {C.~N.}\ \bibnamefont {van derWal}}, \ and\ \bibinfo {author}
  {\bibfnamefont {A.}~\bibnamefont {vanWissen}},\ }\href@noop {} {\bibfield
  {journal} {\bibinfo  {journal} {Auton Agent Multi-Agent Syst}\ ,\ \bibinfo
  {pages} {1}} (\bibinfo {year} {2012})}\BibitemShut {NoStop}%
\bibitem [{\citenamefont {Edelson}\ \emph {et~al.}(2011)\citenamefont
  {Edelson}, \citenamefont {Sharot}, \citenamefont {Dolan},\ and\ \citenamefont
  {Dudai}}]{Edelson2011}%
  \BibitemOpen
  \bibfield  {author} {\bibinfo {author} {\bibfnamefont {M.}~\bibnamefont
  {Edelson}}, \bibinfo {author} {\bibfnamefont {T.}~\bibnamefont {Sharot}},
  \bibinfo {author} {\bibfnamefont {R.~J.}\ \bibnamefont {Dolan}}, \ and\
  \bibinfo {author} {\bibfnamefont {Y.}~\bibnamefont {Dudai}},\ }\href@noop {}
  {\bibfield  {journal} {\bibinfo  {journal} {Science}\ }\textbf {\bibinfo
  {volume} {333}},\ \bibinfo {pages} {108} (\bibinfo {year}
  {2011})}\BibitemShut {NoStop}%
\bibitem [{\citenamefont {Church}\ and\ \citenamefont {Sexton}(2002)}]{CS02}%
  \BibitemOpen
  \bibfield  {author} {\bibinfo {author} {\bibfnamefont {R.~L.}\ \bibnamefont
  {Church}}\ and\ \bibinfo {author} {\bibfnamefont {R.}~\bibnamefont
  {Sexton}},\ }\href@noop {} {\bibfield  {journal} {\bibinfo  {journal}
  {Vehicle Intelligence \& Transportation Analysis Laboratory}\ ,\ \bibinfo
  {pages} {1}} (\bibinfo {year} {2002})}\BibitemShut {NoStop}%
\bibitem [{\citenamefont {of~Transportation}\ and\ \citenamefont
  {of~Homeland~Security}(2006)}]{DOT06}%
  \BibitemOpen
  \bibfield  {author} {\bibinfo {author} {\bibfnamefont {U.~D.}\ \bibnamefont
  {of~Transportation}}\ and\ \bibinfo {author} {\bibfnamefont {U.~D.}\
  \bibnamefont {of~Homeland~Security}},\ }\href@noop {} {\  (\bibinfo {year}
  {2006})}\BibitemShut {NoStop}%
\bibitem [{\citenamefont {Huang}(2011)}]{H11}%
  \BibitemOpen
  \bibfield  {author} {\bibinfo {author} {\bibfnamefont {L.}~\bibnamefont
  {Huang}},\ }\href@noop {} {\bibfield  {journal} {\bibinfo  {journal} {FIU
  Electronic Theses and. Dissertations}\ ,\ \bibinfo {pages} {1}} (\bibinfo
  {year} {2011})}\BibitemShut {NoStop}%
\bibitem [{\citenamefont {Sood}\ \emph {et~al.}(2008)\citenamefont {Sood},
  \citenamefont {Antal},\ and\ \citenamefont {Redner}}]{Sood2008}%
  \BibitemOpen
  \bibfield  {author} {\bibinfo {author} {\bibfnamefont {V.}~\bibnamefont
  {Sood}}, \bibinfo {author} {\bibfnamefont {T.}~\bibnamefont {Antal}}, \ and\
  \bibinfo {author} {\bibfnamefont {S.}~\bibnamefont {Redner}},\ }\href
  {\doibase 10.1103/PhysRevE.77.041121} {\bibfield  {journal} {\bibinfo
  {journal} {Physical Review E}\ }\textbf {\bibinfo {volume} {77}},\ \bibinfo
  {pages} {041121} (\bibinfo {year} {2008})}\BibitemShut {NoStop}%
\bibitem [{\citenamefont {Gigerenzer}(2002)}]{Gigerenzer2002}%
  \BibitemOpen
  \bibfield  {author} {\bibinfo {author} {\bibfnamefont {G.}~\bibnamefont
  {Gigerenzer}},\ }\href@noop {} {\emph {\bibinfo {title} {Calculated Risks:
  How To Know When Numbers Deceive You}}}\ (\bibinfo  {publisher} {Simon \&
  Schuster},\ \bibinfo {year} {2002})\BibitemShut {NoStop}%
\bibitem [{\citenamefont {Chib}\ \emph {et~al.}(2012)\citenamefont {Chib},
  \citenamefont {De~Martino}, \citenamefont {Shimojo},\ and\ \citenamefont
  {O'Doherty}}]{Chib2012}%
  \BibitemOpen
  \bibfield  {author} {\bibinfo {author} {\bibfnamefont {V.}~\bibnamefont
  {Chib}}, \bibinfo {author} {\bibfnamefont {B.}~\bibnamefont {De~Martino}},
  \bibinfo {author} {\bibfnamefont {S.}~\bibnamefont {Shimojo}}, \ and\
  \bibinfo {author} {\bibfnamefont {J.}~\bibnamefont {O'Doherty}},\ }\href@noop
  {} {\bibfield  {journal} {\bibinfo  {journal} {Neuron}\ }\textbf {\bibinfo
  {volume} {74}},\ \bibinfo {pages} {582} (\bibinfo {year} {2012})}\BibitemShut
  {NoStop}%
\bibitem [{\citenamefont {Edwards}\ \emph {et~al.}(2012)\citenamefont
  {Edwards}, \citenamefont {Snyder}, \citenamefont {Allen}, \citenamefont
  {Makinson},\ and\ \citenamefont {Hamby}}]{Edwards2012}%
  \BibitemOpen
  \bibfield  {author} {\bibinfo {author} {\bibfnamefont {J.}~\bibnamefont
  {Edwards}}, \bibinfo {author} {\bibfnamefont {F.}~\bibnamefont {Snyder}},
  \bibinfo {author} {\bibfnamefont {P.}~\bibnamefont {Allen}}, \bibinfo
  {author} {\bibfnamefont {K.}~\bibnamefont {Makinson}}, \ and\ \bibinfo
  {author} {\bibfnamefont {D.}~\bibnamefont {Hamby}},\ }\href {\doibase
  10.1111/j.1539-6924.2012.01839.x.} {\bibfield  {journal} {\bibinfo  {journal}
  {Risk Anal.}\ } (\bibinfo {year} {2012}),\
  10.1111/j.1539-6924.2012.01839.x.}\BibitemShut {Stop}%
\bibitem [{\citenamefont {D.~R.~Bach}(2012)}]{BD12}%
  \BibitemOpen
  \bibfield  {author} {\bibinfo {author} {\bibfnamefont {R.~J.~D.}\
  \bibnamefont {D.~R.~Bach}},\ }\href@noop {} {\bibfield  {journal} {\bibinfo
  {journal} {Nature Reviews Neuroscience}\ }\textbf {\bibinfo {volume} {13}},\
  \bibinfo {pages} {572} (\bibinfo {year} {2012})}\BibitemShut {NoStop}%
\bibitem [{\citenamefont {John}\ \emph {et~al.}(2008)\citenamefont {John},
  \citenamefont {Naumann},\ and\ \citenamefont {Soto}}]{John2008}%
  \BibitemOpen
  \bibfield  {author} {\bibinfo {author} {\bibfnamefont {O.~P.}\ \bibnamefont
  {John}}, \bibinfo {author} {\bibfnamefont {L.~P.}\ \bibnamefont {Naumann}}, \
  and\ \bibinfo {author} {\bibfnamefont {C.~J.}\ \bibnamefont {Soto}},\ }in\
  \href@noop {} {\emph {\bibinfo {booktitle} {Handbook of personality: Theory
  and research}}},\ \bibinfo {editor} {edited by\ \bibinfo {editor}
  {\bibfnamefont {O.~P.}\ \bibnamefont {John}}, \bibinfo {editor}
  {\bibfnamefont {R.~W.}\ \bibnamefont {Robins}}, \ and\ \bibinfo {editor}
  {\bibfnamefont {L.~A.}\ \bibnamefont {Pervin}}}\ (\bibinfo  {publisher}
  {Guilford Press},\ \bibinfo {year} {2008})\ pp.\ \bibinfo {pages}
  {114--158}\BibitemShut {NoStop}%
\bibitem [{\citenamefont {John}\ \emph {et~al.}(1991)\citenamefont {John},
  \citenamefont {Donahue},\ and\ \citenamefont {Kentle}}]{John1991}%
  \BibitemOpen
  \bibfield  {author} {\bibinfo {author} {\bibfnamefont {O.~P.}\ \bibnamefont
  {John}}, \bibinfo {author} {\bibfnamefont {E.~M.}\ \bibnamefont {Donahue}}, \
  and\ \bibinfo {author} {\bibfnamefont {R.~L.}\ \bibnamefont {Kentle}},\
  }\href@noop {} {\enquote {\bibinfo {title} {The big five inventory--versions
  4a and 54},}\ }\bibinfo {howpublished} {Berkeley, CA: University of
  California,Berkeley, Institute of Personality and Social Research} (\bibinfo
  {year} {1991})\BibitemShut {NoStop}%
\bibitem [{\citenamefont {Benet-Martinez}\ and\ \citenamefont
  {John}(1998)}]{Benet1998}%
  \BibitemOpen
  \bibfield  {author} {\bibinfo {author} {\bibfnamefont {V.}~\bibnamefont
  {Benet-Martinez}}\ and\ \bibinfo {author} {\bibfnamefont {O.~P.}\
  \bibnamefont {John}},\ }\href@noop {} {\bibfield  {journal} {\bibinfo
  {journal} {Journal of Personality and Social Psychology}\ }\textbf {\bibinfo
  {volume} {75}},\ \bibinfo {pages} {729} (\bibinfo {year} {1998})}\BibitemShut
  {NoStop}%
\bibitem [{\citenamefont {Weber}\ \emph {et~al.}(2002)\citenamefont {Weber},
  \citenamefont {Blais},\ and\ \citenamefont {Betz}}]{Weber2002}%
  \BibitemOpen
  \bibfield  {author} {\bibinfo {author} {\bibfnamefont {E.~U.}\ \bibnamefont
  {Weber}}, \bibinfo {author} {\bibfnamefont {A.-R.}\ \bibnamefont {Blais}}, \
  and\ \bibinfo {author} {\bibfnamefont {N.}~\bibnamefont {Betz}},\ }\href@noop
  {} {\bibfield  {journal} {\bibinfo  {journal} {Journal of Behavioral Decision
  Making}\ }\textbf {\bibinfo {volume} {15}},\ \bibinfo {pages} {263} (\bibinfo
  {year} {2002})}\BibitemShut {NoStop}%
\bibitem [{\citenamefont {Blais}\ and\ \citenamefont
  {Weber}(2006)}]{Blais2006}%
  \BibitemOpen
  \bibfield  {author} {\bibinfo {author} {\bibfnamefont {A.-R.}\ \bibnamefont
  {Blais}}\ and\ \bibinfo {author} {\bibfnamefont {E.~U.}\ \bibnamefont
  {Weber}},\ }\href@noop {} {\bibfield  {journal} {\bibinfo  {journal}
  {Judgment and Decision Making}\ }\textbf {\bibinfo {volume} {1}},\ \bibinfo
  {pages} {33} (\bibinfo {year} {2006})}\BibitemShut {NoStop}%
\bibitem [{\citenamefont {Srivastava}\ \emph {et~al.}(2003)\citenamefont
  {Srivastava}, \citenamefont {John}, \citenamefont {Gosling},\ and\
  \citenamefont {Potter}}]{Srivastava2003}%
  \BibitemOpen
  \bibfield  {author} {\bibinfo {author} {\bibfnamefont {S.}~\bibnamefont
  {Srivastava}}, \bibinfo {author} {\bibfnamefont {O.~P.}\ \bibnamefont
  {John}}, \bibinfo {author} {\bibfnamefont {S.~D.}\ \bibnamefont {Gosling}}, \
  and\ \bibinfo {author} {\bibfnamefont {J.}~\bibnamefont {Potter}},\
  }\href@noop {} {\bibfield  {journal} {\bibinfo  {journal} {Journal of
  Personality and Social Psychology}\ }\textbf {\bibinfo {volume} {84}},\
  \bibinfo {pages} {1041} (\bibinfo {year} {2003})}\BibitemShut {NoStop}%
\bibitem [{\citenamefont {Bayati}\ \emph {et~al.}(2010)\citenamefont {Bayati},
  \citenamefont {Kim},\ and\ \citenamefont {Saberi}}]{BayatiEtAl2010}%
  \BibitemOpen
  \bibfield  {author} {\bibinfo {author} {\bibfnamefont {M.}~\bibnamefont
  {Bayati}}, \bibinfo {author} {\bibfnamefont {J.~H.}\ \bibnamefont {Kim}}, \
  and\ \bibinfo {author} {\bibfnamefont {A.}~\bibnamefont {Saberi}},\ }\href
  {\doibase 10.1007/s00453-009-9340-1} {\bibfield  {journal} {\bibinfo
  {journal} {Algorithmica}\ }\textbf {\bibinfo {volume} {58}},\ \bibinfo
  {pages} {860} (\bibinfo {year} {2010})}\BibitemShut {NoStop}%
\bibitem [{\citenamefont {Hagberg}\ \emph {et~al.}(2008)\citenamefont
  {Hagberg}, \citenamefont {Schult},\ and\ \citenamefont
  {Swart}}]{hagberg-2008-exploring}%
  \BibitemOpen
  \bibfield  {author} {\bibinfo {author} {\bibfnamefont {A.~A.}\ \bibnamefont
  {Hagberg}}, \bibinfo {author} {\bibfnamefont {D.~A.}\ \bibnamefont {Schult}},
  \ and\ \bibinfo {author} {\bibfnamefont {P.~J.}\ \bibnamefont {Swart}},\ }in\
  \href@noop {} {\emph {\bibinfo {booktitle} {Proceedings of the 7th Python in
  Science Conference (SciPy2008)}}}\ (\bibinfo {address} {Pasadena, CA USA},\
  \bibinfo {year} {2008})\ pp.\ \bibinfo {pages} {11--15}\BibitemShut {NoStop}%
\bibitem [{\citenamefont {Easley}\ and\ \citenamefont
  {Kleinberg}(2010)}]{Easley2010}%
  \BibitemOpen
  \bibfield  {author} {\bibinfo {author} {\bibfnamefont {D.}~\bibnamefont
  {Easley}}\ and\ \bibinfo {author} {\bibfnamefont {J.}~\bibnamefont
  {Kleinberg}},\ }\href@noop {} {\emph {\bibinfo {title} {Networks, Crowds, and
  Markets: {R}easoning About a Highly Connected World}}}\ (\bibinfo
  {publisher} {Cambridge University Press},\ \bibinfo {year}
  {2010})\BibitemShut {NoStop}%
\bibitem [{\citenamefont {Crews}(2012)}]{CrewsThesis}%
  \BibitemOpen
  \bibfield  {author} {\bibinfo {author} {\bibfnamefont {J.}~\bibnamefont
  {Crews}},\ }\emph {\bibinfo {title} {Determining Optimal Evacuation Decision
  Policies for Disasters}},\ \href@noop {} {Master's thesis},\ \bibinfo
  {school} {Naval Postgraduate School}, \bibinfo {address} {Monterey, CA}
  (\bibinfo {year} {2012})\BibitemShut {NoStop}%
\bibitem [{\citenamefont {Soane}\ and\ \citenamefont
  {Chmiel}(2005)}]{Soane2005}%
  \BibitemOpen
  \bibfield  {author} {\bibinfo {author} {\bibfnamefont {E.}~\bibnamefont
  {Soane}}\ and\ \bibinfo {author} {\bibfnamefont {N.}~\bibnamefont {Chmiel}},\
  }\href@noop {} {\bibfield  {journal} {\bibinfo  {journal} {Personality and
  Individual Differences}\ }\textbf {\bibinfo {volume} {38}},\ \bibinfo {pages}
  {1781} (\bibinfo {year} {2005})}\BibitemShut {NoStop}%
\bibitem [{\citenamefont {Otto}\ and\ \citenamefont
  {Day}(2007)}]{otto_biologists_2007}%
  \BibitemOpen
  \bibfield  {author} {\bibinfo {author} {\bibfnamefont {S.~P.}\ \bibnamefont
  {Otto}}\ and\ \bibinfo {author} {\bibfnamefont {T.}~\bibnamefont {Day}},\
  }\href@noop {} {\emph {\bibinfo {title} {A Biologist's Guide to Mathematical
  Modeling in Ecology and Evolution}}}\ (\bibinfo  {publisher} {Princeton
  University Press},\ \bibinfo {year} {2007})\BibitemShut {NoStop}%
\bibitem [{\citenamefont {Shuler}\ and\ \citenamefont
  {Kargi}(2001)}]{shuler_bioprocess_2001}%
  \BibitemOpen
  \bibfield  {author} {\bibinfo {author} {\bibfnamefont {M.~L.}\ \bibnamefont
  {Shuler}}\ and\ \bibinfo {author} {\bibfnamefont {F.}~\bibnamefont {Kargi}},\
  }\href@noop {} {\emph {\bibinfo {title} {Bioprocess Engineering: Basic
  Concepts}}},\ \bibinfo {edition} {2nd}\ ed.\ (\bibinfo  {publisher} {Prentice
  Hall},\ \bibinfo {year} {2001})\BibitemShut {NoStop}%
\bibitem [{\citenamefont {Granovetter}(1978)}]{Granovetter1978}%
  \BibitemOpen
  \bibfield  {author} {\bibinfo {author} {\bibfnamefont {M.}~\bibnamefont
  {Granovetter}},\ }\href@noop {} {\bibfield  {journal} {\bibinfo  {journal}
  {The American Journal of Sociology}\ }\textbf {\bibinfo {volume} {83}},\
  \bibinfo {pages} {1420} (\bibinfo {year} {1978})}\BibitemShut {NoStop}%
\bibitem [{\citenamefont {Macy}(1991)}]{macy_chains_1991}%
  \BibitemOpen
  \bibfield  {author} {\bibinfo {author} {\bibfnamefont {M.~W.}\ \bibnamefont
  {Macy}},\ }\href@noop {} {\bibfield  {journal} {\bibinfo  {journal} {American
  Sociological Review}\ }\textbf {\bibinfo {volume} {56}},\ \bibinfo {pages}
  {730} (\bibinfo {year} {1991})}\BibitemShut {NoStop}%
\bibitem [{\citenamefont {Bevington}\ and\ \citenamefont
  {Robinson}(2002)}]{bevington_data_2002}%
  \BibitemOpen
  \bibfield  {author} {\bibinfo {author} {\bibfnamefont {P.}~\bibnamefont
  {Bevington}}\ and\ \bibinfo {author} {\bibfnamefont {D.~K.}\ \bibnamefont
  {Robinson}},\ }\href@noop {} {\emph {\bibinfo {title} {Data Reduction and
  Error Analysis for the Physical Sciences}}},\ \bibinfo {edition} {3rd}\ ed.\
  (\bibinfo  {publisher} {{McGraw-Hill} {Science/Engineering/Math}},\ \bibinfo
  {year} {2002})\BibitemShut {NoStop}%
\bibitem [{\citenamefont {Press}\ \emph {et~al.}(1992)\citenamefont {Press},
  \citenamefont {Flannery}, \citenamefont {Teukolsky},\ and\ \citenamefont
  {Vetterling}}]{press_numerical_1992}%
  \BibitemOpen
  \bibfield  {author} {\bibinfo {author} {\bibfnamefont {W.~H.}\ \bibnamefont
  {Press}}, \bibinfo {author} {\bibfnamefont {B.~P.}\ \bibnamefont {Flannery}},
  \bibinfo {author} {\bibfnamefont {S.~A.}\ \bibnamefont {Teukolsky}}, \ and\
  \bibinfo {author} {\bibfnamefont {W.~T.}\ \bibnamefont {Vetterling}},\
  }\href@noop {} {\emph {\bibinfo {title} {Numerical Recipes in C: The Art of
  Scientific Computing, Second Edition}}},\ \bibinfo {edition} {2nd}\ ed.\
  (\bibinfo  {publisher} {Cambridge University Press},\ \bibinfo {year}
  {1992})\BibitemShut {NoStop}%
\bibitem [{\citenamefont {Hastie}\ \emph {et~al.}(2009)\citenamefont {Hastie},
  \citenamefont {Tibshirani},\ and\ \citenamefont
  {Friedman}}]{hastie_elements_2009}%
  \BibitemOpen
  \bibfield  {author} {\bibinfo {author} {\bibfnamefont {T.}~\bibnamefont
  {Hastie}}, \bibinfo {author} {\bibfnamefont {R.}~\bibnamefont {Tibshirani}},
  \ and\ \bibinfo {author} {\bibfnamefont {J.}~\bibnamefont {Friedman}},\
  }\href@noop {} {\emph {\bibinfo {title} {The elements of statistical learning
  data mining, inference, and prediction}}}\ (\bibinfo  {publisher}
  {Springer},\ \bibinfo {address} {New York},\ \bibinfo {year}
  {2009})\BibitemShut {NoStop}%
\bibitem [{\citenamefont {McCaffrey}(2004)}]{McCaffrey2004}%
  \BibitemOpen
  \bibfield  {author} {\bibinfo {author} {\bibfnamefont {S.}~\bibnamefont
  {McCaffrey}},\ }\href@noop {} {\bibfield  {journal} {\bibinfo  {journal}
  {Society and Natural Resources}\ }\textbf {\bibinfo {volume} {17}},\ \bibinfo
  {pages} {509} (\bibinfo {year} {2004})}\BibitemShut {NoStop}%
\bibitem [{\citenamefont {Regnier}(2008{\natexlab{a}})}]{R08}%
  \BibitemOpen
  \bibfield  {author} {\bibinfo {author} {\bibfnamefont {E.}~\bibnamefont
  {Regnier}},\ }\href@noop {} {\bibfield  {journal} {\bibinfo  {journal}
  {Management Science}\ }\textbf {\bibinfo {volume} {54}},\ \bibinfo {pages}
  {16Ð28} (\bibinfo {year} {2008}{\natexlab{a}})}\BibitemShut {NoStop}%
\bibitem [{\citenamefont {Service}(2012)}]{NOAA.GOV}%
  \BibitemOpen
  \bibfield  {author} {\bibinfo {author} {\bibfnamefont {N.~W.}\ \bibnamefont
  {Service}},\ }\href
  {http://www.nhc.noaa.gov/archive/2012/graphics/al18/loop_PROB34.shtml}
  {\enquote {\bibinfo {title} {Sandy graphics archive {@ONLINE}},}\ } (\bibinfo
  {year} {2012})\BibitemShut {NoStop}%
\bibitem [{\citenamefont {Regnier}(2008{\natexlab{b}})}]{regnier2008}%
  \BibitemOpen
  \bibfield  {author} {\bibinfo {author} {\bibfnamefont {E.}~\bibnamefont
  {Regnier}},\ }\href@noop {} {\bibfield  {journal} {\bibinfo  {journal}
  {Management Science}\ }\textbf {\bibinfo {volume} {54}},\ \bibinfo {pages}
  {16} (\bibinfo {year} {2008}{\natexlab{b}})}\BibitemShut {NoStop}%
\bibitem [{\citenamefont {Dabner}(2012)}]{Dabner2012}%
  \BibitemOpen
  \bibfield  {author} {\bibinfo {author} {\bibfnamefont {N.}~\bibnamefont
  {Dabner}},\ }\href@noop {} {\bibfield  {journal} {\bibinfo  {journal} {The
  Internet and Higher Education}\ }\textbf {\bibinfo {volume} {15}},\ \bibinfo
  {pages} {69} (\bibinfo {year} {2012})}\BibitemShut {NoStop}%
\bibitem [{\citenamefont {Oswald}()}]{Oswald2011}%
  \BibitemOpen
  \bibfield  {author} {\bibinfo {author} {\bibfnamefont {E.}~\bibnamefont
  {Oswald}},\ }\href@noop {} {\enquote {\bibinfo {title} {{New Yorkers saw DC
  quake tweets before the ground shook}},}\ }\bibinfo {howpublished}
  {\url{http://betanews.com/2011/08/23/new-yorkers-saw-dc-quake-tweets-before-the-ground-shook/}},\
  \bibinfo {note} {published: 23 August 2011. Last accessed: 2 January
  2012.}\BibitemShut {Stop}%
\bibitem [{\citenamefont {Ball}()}]{Ball2011}%
  \BibitemOpen
  \bibfield  {author} {\bibinfo {author} {\bibfnamefont {D.}~\bibnamefont
  {Ball}},\ }\href@noop {} {\enquote {\bibinfo {title} {Hurricane, earthquake
  show utility of social media},}\ }\bibinfo {howpublished}
  {\url{http://www.forbes.com/sites/gyro/2011/08/31/hurricane-earthquake-show-utility-of-social-media/}},\
  \bibinfo {note} {published: 31 August 2011. Last accessed: 2 January
  2012.}\BibitemShut {Stop}%
\bibitem [{\citenamefont {Behrens}\ \emph {et~al.}(2009)\citenamefont
  {Behrens}, \citenamefont {Hunt},\ and\ \citenamefont
  {Rushworth}}]{Behrens2009}%
  \BibitemOpen
  \bibfield  {author} {\bibinfo {author} {\bibfnamefont {T.~E.}\ \bibnamefont
  {Behrens}}, \bibinfo {author} {\bibfnamefont {L.~T.}\ \bibnamefont {Hunt}}, \
  and\ \bibinfo {author} {\bibfnamefont {M.~F.}\ \bibnamefont {Rushworth}},\
  }\href@noop {} {\bibfield  {journal} {\bibinfo  {journal} {Science}\ }\textbf
  {\bibinfo {volume} {324}},\ \bibinfo {pages} {1160} (\bibinfo {year}
  {2009})}\BibitemShut {NoStop}%
\bibitem [{\citenamefont {Glimcher}(2004)}]{Glimcher2004}%
  \BibitemOpen
  \bibfield  {author} {\bibinfo {author} {\bibfnamefont {P.~W.}\ \bibnamefont
  {Glimcher}},\ }\href@noop {} {\emph {\bibinfo {title} {Decisions,
  Uncertainty, and the Brain: The Science of Neuroeconomics}}}\ (\bibinfo
  {publisher} {Bradford Books},\ \bibinfo {year} {2004})\BibitemShut {NoStop}%
\bibitem [{\citenamefont {Rangel}\ \emph {et~al.}(2008)\citenamefont {Rangel},
  \citenamefont {Camerer},\ and\ \citenamefont {Montague}}]{Rangel2008}%
  \BibitemOpen
  \bibfield  {author} {\bibinfo {author} {\bibfnamefont {A.}~\bibnamefont
  {Rangel}}, \bibinfo {author} {\bibfnamefont {C.}~\bibnamefont {Camerer}}, \
  and\ \bibinfo {author} {\bibfnamefont {P.~R.}\ \bibnamefont {Montague}},\
  }\href@noop {} {\bibfield  {journal} {\bibinfo  {journal} {Nat Rev Neurosci}\
  }\textbf {\bibinfo {volume} {9}},\ \bibinfo {pages} {545} (\bibinfo {year}
  {2008})}\BibitemShut {NoStop}%
\bibitem [{\citenamefont {Venkatraman}\ \emph {et~al.}(2009)\citenamefont
  {Venkatraman}, \citenamefont {Payne}, \citenamefont {Bettman}, \citenamefont
  {Luce},\ and\ \citenamefont {Huettel}}]{Venkatraman2009}%
  \BibitemOpen
  \bibfield  {author} {\bibinfo {author} {\bibfnamefont {V.}~\bibnamefont
  {Venkatraman}}, \bibinfo {author} {\bibfnamefont {J.~W.}\ \bibnamefont
  {Payne}}, \bibinfo {author} {\bibfnamefont {J.~R.}\ \bibnamefont {Bettman}},
  \bibinfo {author} {\bibfnamefont {M.~F.}\ \bibnamefont {Luce}}, \ and\
  \bibinfo {author} {\bibfnamefont {S.~A.}\ \bibnamefont {Huettel}},\
  }\href@noop {} {\bibfield  {journal} {\bibinfo  {journal} {Neuron}\ }\textbf
  {\bibinfo {volume} {62}},\ \bibinfo {pages} {593} (\bibinfo {year}
  {2009})}\BibitemShut {NoStop}%
\bibitem [{\citenamefont {Kolling}\ \emph {et~al.}(2012)\citenamefont
  {Kolling}, \citenamefont {Behrens}, \citenamefont {Mars},\ and\ \citenamefont
  {Rushworth}}]{Kolling2012}%
  \BibitemOpen
  \bibfield  {author} {\bibinfo {author} {\bibfnamefont {N.}~\bibnamefont
  {Kolling}}, \bibinfo {author} {\bibfnamefont {T.~E.}\ \bibnamefont
  {Behrens}}, \bibinfo {author} {\bibfnamefont {R.~B.}\ \bibnamefont {Mars}}, \
  and\ \bibinfo {author} {\bibfnamefont {M.~F.}\ \bibnamefont {Rushworth}},\
  }\href@noop {} {\bibfield  {journal} {\bibinfo  {journal} {Science}\ }\textbf
  {\bibinfo {volume} {336}},\ \bibinfo {pages} {95} (\bibinfo {year}
  {2012})}\BibitemShut {NoStop}%
\bibitem [{\citenamefont {Chang}\ \emph {et~al.}(2012)\citenamefont {Chang},
  \citenamefont {Barack},\ and\ \citenamefont {Platt}}]{Chang2012}%
  \BibitemOpen
  \bibfield  {author} {\bibinfo {author} {\bibfnamefont {S.~W.}\ \bibnamefont
  {Chang}}, \bibinfo {author} {\bibfnamefont {D.~L.}\ \bibnamefont {Barack}}, \
  and\ \bibinfo {author} {\bibfnamefont {M.~L.}\ \bibnamefont {Platt}},\
  }\href@noop {} {\bibfield  {journal} {\bibinfo  {journal} {Biol Psychiatry}\
  }\textbf {\bibinfo {volume} {72}},\ \bibinfo {pages} {101} (\bibinfo {year}
  {2012})}\BibitemShut {NoStop}%
\bibitem [{\citenamefont {Hartley}\ and\ \citenamefont
  {Phelps}(2012)}]{Hartley2012}%
  \BibitemOpen
  \bibfield  {author} {\bibinfo {author} {\bibfnamefont {C.~A.}\ \bibnamefont
  {Hartley}}\ and\ \bibinfo {author} {\bibfnamefont {E.~A.}\ \bibnamefont
  {Phelps}},\ }\href@noop {} {\bibfield  {journal} {\bibinfo  {journal} {Biol
  Psychiatry}\ }\textbf {\bibinfo {volume} {72}},\ \bibinfo {pages} {113}
  (\bibinfo {year} {2012})}\BibitemShut {NoStop}%
\end{thebibliography}%


\end{document}